\documentclass[11pt,a4paper]{article}

\usepackage[english]{babel}
\usepackage{amsmath}
\usepackage{amssymb,amsfonts,a4}
\usepackage{eucal,bbm,stmaryrd}
\usepackage{tikz}
\usetikzlibrary{patterns}

\begin{document}

\numberwithin{equation}{section}

\newtheorem{satz}{Theorem}
\newtheorem{defi}{Definition}
\newtheorem{lem}{Lemma}
\newtheorem{kor}{Corollary}

%Vorbereitungen

\newcommand{\Bew}{\noindent{\bf Proof: }}
\newcommand{\qed}{\hspace*{3em} \hfill{$\square$}}
\newcommand{\weg}{\setminus}
\newcommand{\gdw}{\Leftrightarrow}
\newcommand{\sgdw}{\Leftrightarrow}
\newcommand{\nach}{\longrightarrow}
\newcommand{\alle}{\,\forall\,}
\newcommand{\gibt}{\,\exists\,}
\newcommand{\Bed}{\,|\,}

\newcommand{\Z}{\mathbbm{Z}}
\newcommand{\R}{\mathbbm{R}}
\newcommand{\baR}{\overline{\R}}
\newcommand{\N}{\mathbbm{N}}

\newcommand{\bxi}{\bar{\xi}}

\newcommand{\X}{\mathcal{X}}
\newcommand{\XX}{\mathcal{X}^{h}}
\newcommand{\bX}{Y}
\newcommand{\La}{\Lambda}
\newcommand{\Lan}{\La_n}

\newcommand{\Bo}{\mathcal{B}}
\newcommand{\B}{\mathcal{B}}
\newcommand{\F}{\mathcal{F}}

\newcommand{\la}{\lambda}
\newcommand{\si}{\sigma}

\newcommand{\einh}{e_1}

%Hauptteil

\newcommand{\Gn}{G_n}

\newcommand{\tn}{\tau_n}

\newcommand{\ho}{\circ}

\newcommand{\mxt}{m_{p,t}}
\newcommand{\hxt}{h_{p,t}}

\newcommand{\de}{\delta}
\newcommand{\ep}{\epsilon}

\newcommand{\ph}{\varphi_n}
\newcommand{\bph}{\bar{\varphi}_n}
\newcommand{\leb}{\lambda^2}

\newcommand{\per}{\eta}

\newcommand{\T}{\mathfrak{T}}
\newcommand{\Tn}{\T_n}
\newcommand{\Tnx}{\T_{n,X}}

\newcommand{\etn}{t_{n}}
\newcommand{\eTn}{T_{n}}
\newcommand{\etaun}{\tau_{n}}
\newcommand{\epn}{P_n}
\newcommand{\ecn}{C_n}

\newcommand{\tX}{\tilde{X}}

\newcommand{\tTn}{\tilde{\T}_n}
\newcommand{\tTnx}{\tilde{\T}_{n,\tX}}

\newcommand{\ettn}{\tilde{t}_{n}}
\newcommand{\etTn}{\tilde{T}_{n}}
\newcommand{\ettaun}{\tilde{\tau}_{n}}
\newcommand{\etpn}{\tilde{P}_n}
\newcommand{\etcn}{\tilde{C}_n}

\newcommand{\iTn}{\bar{\T}_n}
\newcommand{\iTnx}{\bar{\T}_{n,X}}

\newcommand{\anx}{a_{n,X}}
\newcommand{\rnx}{r_{n,X}}

\newcommand{\tnx}{t_{n,X}}
\newcommand{\taunx}{\tau_{n,X}}
\newcommand{\pnx}{P_{n,X}}
\newcommand{\Tnxn}{T_{n,X}}

\newcommand{\ttnx}{\tilde{t}_{n,\tX}}
\newcommand{\tTnxn}{\tilde{T}_{n,\tX}}
\newcommand{\ttaunx}{\tilde{\tau}_{n,\tX}}
\newcommand{\tpnx}{\tilde{P}_{n,\tX}}
\newcommand{\ak}{A_{\per}}
\newcommand{\tak}{\tilde{A}_{\per}}

%========================================================================

{\bf \Large
\begin{center}
Lower bound on the mean square displacement\\
of particles in the hard disk model
\end{center}}

\vspace{0 cm}

\begin{center}
Thomas Richthammer\\
IMAI, Universit\"at Hildesheim\\
Samelsonplatz 1, D-31141 Hildesheim\\
Email: richthammer@imai.uni-hildesheim.de\\
Tel: +49 5121 883 40160
\end{center}

\vspace{0 cm}

\begin{abstract}
The hard disk model is a 2D Gibbsian process of particles 
interacting via pure hard core repulsion. At high particle density 
the model is believed to show orientational order, however, 
it is known not to exhibit positional order. 
Here we investigate to what extent particle positions may fluctuate. 
We consider a finite volume version of the model 
in a box of dimensions $2n \times 2n$ with arbitrary boundary configuration,
and we show that the mean square displacement of particles near the center of the box
is bounded from below by $c \log n$. 
The result generalizes to a large class of models with fairly arbitrary interaction. 
\\
 
\end{abstract}

%=========================================================================

\section{Introduction}

\begin{sloppypar}
Crystallization is an important topic in equilibrium state statistical physics. 
At low temperature or at high particle density particles arrange themselves 
into regular patterns such as lattice-like structures and thus form a state 
which is usually referred to as solid. It is not always clear to which extent 
such a solid indeed resembles a lattice. In particular for 2D solids this has been 
debated for a long time. A first model to be studied was the 2D harmonic 
crystal. Peierls showed that in this model particles are not localized (see 
\cite{P1}, \cite{P2}), i.e. with increasing size of the system fluctuations 
of particle positions grow unboundedly. More precisely the mean square 
displacement of a particle from its ideal lattice position is of order $\log n$ 
if $n$ is the size of the system. The absence of positional order was 
shown to be a general feature of 2D particle systems. This follows from ideas 
of Mermin and Wagner (\cite{MW} and \cite{M}) and was first applied to 
particle positions in a continuum setting by Fr\"ohlich and Pfister (\cite{FP1} 
and \cite{FP2}). In spite of this negative result, 2D solids are believed to exist, 
and indeed their lattice-like structure may be due to orientational order 
rather than positional order. So far this could not be shown rigorously 
for any realistic particle system, but this is supported by results obtained 
from simulations (e.g. see \cite{BK}) and results in related models with a predefined 
lattice structure used to label particles (see \cite{MR}, \cite{HMR}, 
\cite{G} and \cite{Au}). 

A particularly simple model with the above properties is the hard disk model. 
Here the interaction between particles is a pure hard-core repulsion, i.e. 
any two point particles are forced to keep a distance of $>2r$ but do not interact 
otherwise (see \cite{L} for a review of properties of this model). 
Equivalently particles can be thought of as disks with radius $r$ 
and the interaction prevents disks from overlapping. 
Besides $r$ the only parameter of the model is the activity $z$ 
regulating the particle density. The model can be obtained from a Poisson point process 
with intensity $z$ by conditioning on particles to keep distance $> 2r$.  
Due to its simplicity it is a good starting point for 
investigations, and it has been studied extensively through simulations. 
According to these simulations the model exhibits three different phases 
(see \cite{BK}): 
\begin{itemize}
\item small $z$: liquid (or gas) phase. The model does not exhibit any order
and we have exponential decay of positional and orientational correlations.
\item intermediate $z$: hexatic phase. The model exhibits orientational order 
and exponential decay of positional correlations. 
\item large $z$: solid phase. The model exhibits orientational order and 
algebraic decay of positional correlations. 
\end{itemize}
The phase transition from liquid to hexatic is of first order
and the phase transition from hexatic to solid is continuous. 
In contrast to this detailed picture, not much is known rigorously: 
\begin{itemize}
\item
In \cite{R} the result of Fr\"ohlich and Pfister is extended to the 
hard disk model, i.e. it is shown that there is no 
positional order for any value of $z$. 
\item 
In \cite{Ar} a percolation result is obtained:
Suppose that for given $\ep > 0$ any two disks of distance $\le \ep$ are 
connected. It should be expected that in the high density regime (i.e. 
for $z$ sufficiently large) we have percolation of connected disks. 
This is established for $\ep > r$. 
\end{itemize}
In the result presented in this paper we refine the result of \cite{R} 
and show that the hard disk model shows the same behaviour as the 
harmonic crystal, in that the mean square displacement of a particle 
from its ideal lattice position is at least of order $\log n$  if $n$ is the size of the 
system. The formulation of such a result is not 
straightforward: Unlike in the harmonic crystal, 
in the hard disk model there is no a priori labelling of particles that would allow 
to pinpoint a specific particle and investigate the fluctuations of its position.  
Instead, we will describe the fluctuations of positions in terms of a certain transformation of particle configurations within a box of size $2n \times 2n$ 
with the following properties: 
\begin{itemize}
\item Particles near the center of the box are translated by an amount of order
$\sqrt{\log n}$, whereas particles near the boundary are kept fixed. 
\item Locally the transformation almost preserves the relative position of particles. 
In particular the hard core condition is not violated. 
\item The transformation only has a mild impact on the probability measure describing 
the hard disk model. 
\end{itemize}
Our main theorem shows that a transformation with these properties can be 
constructed for the hard disk model, and by means of a corollary 
we will explain why this transformation 
should be thought of as providing a lower bound of order $\log n$  
for the mean square displacement of particles. We note that our results 
are not restricted to pure hard core repulsion but can be extended to 
fairly arbitrary interactions. 

A variant of the transformation described above was the main tool 
used in \cite{R} for showing the absence of positional order. In 
\cite{MP} the method was adjusted to a lattice setting and used to show 
a delocalization result for the random Lipschitz surface model, including 
a lower bound on fluctuations of order $\log n$. We use 
improvements and refinements of the method from \cite{MP} and 
adjust them back to the continuous setting of the hard disk model. 
Some arguments are taken straight from \cite{R} but repeated here for 
the sake of completeness.

In Section \ref{secresult} we give a precise description of our result (theorem 
and corollary) and we outline to what extent it can be generalized. 
In Section \ref{secproofcor} we give a proof of the corollary. 
In Section \ref{secproofthm} we give a proof of the theorem. 
All technical parts and lemmas used in this proof are relegated to 
Section \ref{seclemmas}.

\bigskip

%=====================================================================
%
%=====================================================================

\section{Result} \label{secresult}

Before explaining the hard disk model we describe the general setting. 
On the single particle state space $\R^2$ we consider the Borel-$\si$-algebra 
denoted by $\B^2$ and the Lebesgue-measure denoted by $\la^2$. 
When integrating with respect to $\la^2$ we use the usual abbreviation
$dx := d\la^2(x)$. Our set of particle configurations is 
$$
\X := \{ X \subset \R^2: \# X_\La < \infty
\text{ for every bounded } \La \subset \R^2\},
$$
the set of all locally finite subsets of the plane. Here $X_\La := X \cap \La$ 
denotes the restriction of a configuration $X$ to a set $\La$ and 
$\#$ denotes the cardinality of a set. 
Let $\X_\La := \{X \in \X: X \subset \La\}$ denote the set of all configurations 
in $\La \subset \R$. The $\si$-algebras $\F$ and $\F_\La$ on 
$\X$ and $\X_\La$ respectively are defined to be generated by the 
counting variables $N_{\La'}$ ($\La' \in \B^2$), where 
$N_{\La'}(X) := \# X_{\La'}$. Our reference measure on 
$(\X_\La,\F_\La)$ for a bounded set $\La \in \B^2$ is the distribution 
$\nu_\La$ of the Poisson point process. We have 
$$
\int \nu_\La(dX) f(X)  \, = \, e^{- \leb(\La)} \, \sum_{ k \ge 0} \,  
 \frac{1}{k!} \, \int_{{\La}^k} dx_1 \ldots dx_k \, 
 \,  f(\{x_i : 1 \le i  \le k\} ), 
$$
for any $\F_{\La}$-measurable function $f: \X_\La \to \R_+$. For convenience 
we would like to include the boundary configuration into the reference measure. 
Here any configuration $Y \in \X$ can serve as a boundary configuration, and 
the reference measure with this boundary configuration is denoted by 
$\nu_\La(.|Y)$. It should be thought of producing a Poisson point process inside
$\La$ and the deterministic configuration $Y_{\La^c}$ outside $\La$, i.e. 
$$
\int \nu_{\La}(dX|Y) f(X) \;
= \; \int \nu_\La (dX) f(X_{\La} \cup Y_{\La^c})
$$
for any $\F$-measurable function $f: \X \to \R_+$. 
$\nu_{\La}(.|Y)$ can be considered as a probability measure on $(\X,\F)$ 
or as a probability measure on $(\X_\La,\F_\La)$. 

For the definition of the hard disk model we need to take the hard core 
into account. By rescaling we may assume that the disk diameter equals 1. 
The hard core can be built into the setting by restricting the configuration 
space to the set of hard core configurations 
$$
\XX := \{X \subset \R^2: \forall x,y \in X: x \neq y \Rightarrow |x-y|_2 > 1\}, 
$$
where $|.|_2$ denotes Euclidean distance. Alternatively this hard core repulsion 
can be modelled by the two-body interaction $U: (\R^2)^2 \to \R \cup \{\infty\}$: 
$$
U (x,y) \, := \, 
\left \{
\begin{array}{ll}
\infty \quad &\text{for } \quad  |x-y|_2 \le 1 \\
0 \quad &\text{for } \quad  |x-y|_2 > 1. 
\end{array}
\right.
$$
The a priory particle density is modelled by an activity parameter  $z>0$ 
corresponding to the chemical potential $-\log z$ of the system.
We note that the inverse temperature, which usually serves as a second 
parameter for a model of this type, does not play a role in case
of pure hard core repulsion.  
The hard disk model can now be described in terms of the 
finite volume Gibbs distributions
$\mu_{\La}^z(.|\bX)$ in volume $\La \in \B^2$ (bounded) 
with respect to the boundary configuration $\bX \in \XX$.  
$\mu_\La^z(.|\bX)$ is a probability measure defined by 
$$
\mu_\La^z(dX|\bX) = \frac 1 {Z_\La^z(\bX)} e^{-H_\La(X)} z^{\#X_{\La}} \nu_\La(dX|\bX)
= \frac 1 {Z_\La^z(\bX)} 1_{\XX}(X) z^{\#X_{\La}} \nu_\La(dX|\bX). 
$$
Here 
$$
H_\La(X) := \frac 1 2 \sum_{x \neq y \in X_\La} U(x,y) +  
\sum_{x \in X_\La, y \in X_{\La^c}} U(x,y)
$$ 
denotes the Hamiltonian and 
$Z_\La^z(\bX)$ denotes the partition function which plays the role of 
a normalizing constant. $\mu_\La^z(.|\bX)$ can be interpreted as a Poisson 
point process in $\La$ with intensity $z$ conditioned on the event 
that any two points in $\La$ keep a distance of $\ge 1$ and any point in $\La$ 
keeps a distance of $\ge 1$ to $Y_{\La^c}$. In this paper we only work in 
finite volume, but we would like to note that the model can also be extended to 
infinite volume by means of the Dobrushin-Lanford-Ruelle (DLR) equation: 
Any probability measure $\mu$ on $(\X,\F)$ that is compatible 
with the above finite volume Gibbs distributions for arbitrary $\La$ and $\bX$ 
is called an infinite volume Gibbs measure at activity $z$.

Our aim is to investigate the extent to which particles in typical configurations 
produced by $\mu_\La^z(.|\bX)$ deviate from a lattice structure
in terms of positional order. 
For sake of simplicity we only consider domains of the 
form 
$$
\La_n \,:= \,[-n,n]^2 \, \subset \, \R^2 \qquad (n \in \N).
$$
Consequently we will use the abbreviations $\mu_n^z := \mu_{\Lan}^z$, 
$Z_n^z := Z_{\Lan}^z$, etc. 
The unit vector $e \in \R^2$ will be used for modelling the direction of 
the proposed deviation. For definiteness we will only consider the direction
$$
e := \einh = (1,0).
$$ 
\newpage
Our main result is the following: 
\begin{satz} \label{thmtransformation}
Let  $z > 0$, $\ep := \min\{\frac 1 {48z},\frac 1 4\}$, $\de \in (0,\frac 1 2]$ and $n  \ge 200$.
There is a transformation $\Tn: \X \to \X$ of the form $\Tn(X) = \{x +  \tnx(x)\einh: x \in X\}$ 
with $\tnx: \R^2 \to [0,\infty)$ and there is a set of good configurations $\Gn \in \F$ such that: 
\begin{itemize}
\item[(1)] For all $X \in \X$ and $x \notin \Lan$ we have $\tnx(x) = 0$. 
\item[(2)] For all $X \in \Gn$ and $x \in X_{\La_{\sqrt n}}$ we have  
$\tnx(x) = \de \ep \sqrt{\log n}$. 
\item[(3)] For all $X \in \X$ and $x,y \in X$ we have   
$|\tnx(x)- \tnx(y)| \le \de |x-y|_2$.
\item[(4)] For all $\bX \in \XX$ we have $\mu_n^z(\Gn^c|Y) \le \frac 1 {n}$. 
\item[(5)] $\Tn$ is bijective and $\Tn(\XX) = \XX$.  
\item[(6)] For every $\bX \in \XX$ $\mu_n^z(.|\bX)$ has a density $\ph: \X \to \R$ w.r.t. $\mu_n^z(.|\bX) \circ \Tn^{-1}$.  
\item[(7)] For every $\bX \in \XX$ we have
$\mu_n^z(|\log(\ph \bph)| | \bX) \le 120 \de^2$. 
\end{itemize}
In addition, the transformation $\iTn: \X \to \X$, $\iTn(X) = \{x -  \tnx(x)\einh: x \in X\}$
has properties analogous to (5) and (6). The function $\bph$ appearing in (7) denotes 
the corresponding density. 
\end{satz}   

\bigskip

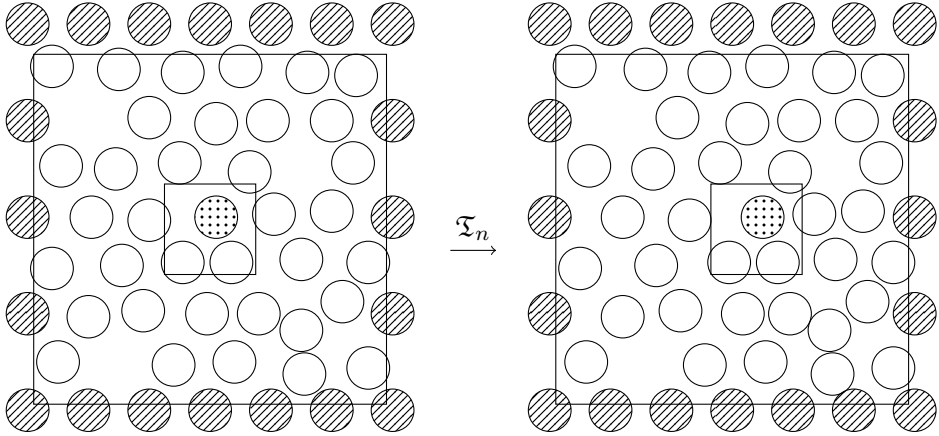
\begin{figure}[!htb] 
\begin{center}
\begin{tikzpicture}[scale = 0.4]
\draw (0.2,0.2) rectangle (11.8,11.8);
\draw (4.5,4.5) rectangle (7.5,7.5);
\draw[pattern = north east lines] (0,0) circle (20 pt); 
\draw[pattern = north east lines] (2,0) circle (20 pt); 
\draw[pattern = north east lines] (4,0) circle (20 pt); 
\draw[pattern = north east lines] (6,0) circle (20 pt); 
\draw[pattern = north east lines] (8,0) circle (20 pt); 
\draw[pattern = north east lines] (10,0) circle (20 pt); 
\draw[pattern = north east lines] (12,0) circle (20 pt); 
\draw (1,1.6) circle (20 pt); 
%\draw (3,1.6) circle (20 pt); 
\draw (4.8,1.5) circle (20 pt); 
\draw (6.8,1.6) circle (20 pt); 
\draw (9.1,1.2) circle (20 pt); 
\draw (11.2,1.4) circle (20 pt); 
\draw[pattern = north east lines] (0,3.2) circle (20 pt); 
\draw (2,3.1) circle (20 pt); 
\draw (3.8,3.3) circle (20 pt); 
\draw (5.9,3.2) circle (20 pt); 
\draw (7.6,3.2) circle (20 pt); 
\draw (10.35,3.6) circle (20 pt); 
\draw (9,2.65) circle (20 pt); 
\draw[pattern = north east lines] (12,3.2) circle (20 pt); 
\draw (0.8,4.7) circle (20 pt); 
\draw (3.1,4.8) circle (20 pt); 
\draw (5.1,4.9) circle (20 pt); 
\draw (6.7,4.9) circle (20 pt); 
\draw (9,5) circle (20 pt); 
\draw (11.2,4.9) circle (20 pt); 
\draw[pattern = north east lines] (0,6.4) circle (20 pt); 
\draw (2.1,6.4) circle (20 pt); 
\draw (4,6.3) circle (20 pt); 
\draw[pattern = dots] (6.2,6.4) circle (20 pt); 
\draw (8.1,6.5) circle (20 pt); 
\draw (10,6.6) circle (20 pt); 
\draw[pattern = north east lines] (12,6.4) circle (20 pt); 
\draw (1.1,8.1) circle (20 pt); 
\draw (2.9,8) circle (20 pt); 
\draw (5,8.2) circle (20 pt); 
\draw (7.3,7.9) circle (20 pt); 
%\draw (9,8) circle (20 pt); 
\draw (10.7,8.2) circle (20 pt); 
\draw[pattern = north east lines] (0,9.6) circle (20 pt); 
%\draw (2,9.6) circle (20 pt); 
\draw (4,9.7) circle (20 pt); 
\draw (6.2,9.5) circle (20 pt); 
\draw (7.9,9.6) circle (20 pt); 
\draw (10,9.6) circle (20 pt); 
\draw[pattern = north east lines] (12,9.6) circle (20 pt); 
\draw (0.8,11.4) circle (20 pt); 
\draw (3,11.3) circle (20 pt); 
\draw (5.1,11.2) circle (20 pt); 
\draw (7,11.4) circle (20 pt); 
\draw (9.2,11.2) circle (20 pt); 
\draw (10.8,11.1) circle (20 pt); 
\draw[pattern = north east lines] (0,12.8) circle (20 pt); 
\draw[pattern = north east lines] (2,12.8) circle (20 pt); 
\draw[pattern = north east lines] (4,12.8) circle (20 pt); 
\draw[pattern = north east lines] (6,12.8) circle (20 pt); 
\draw[pattern = north east lines] (8,12.8) circle (20 pt); 
\draw[pattern = north east lines] (10,12.8) circle (20 pt); 
\draw[pattern = north east lines] (12,12.8) circle (20 pt); 
%\draw[color=red] (2,1.65) circle (11 pt); 
\draw[->](13.9,5.3)-- node[above]{$\Tn$}(15.4,5.3);
\end{tikzpicture}
\;
\begin{tikzpicture}[scale = 0.4]
\draw (0.2,0.2) rectangle (11.8,11.8);
\draw (5.3,4.5) rectangle (8.3,7.5);
\draw[pattern = north east lines] (0,0) circle (20 pt); 
\draw[pattern = north east lines] (2,0) circle (20 pt); 
\draw[pattern = north east lines] (4,0) circle (20 pt); 
\draw[pattern = north east lines] (6,0) circle (20 pt); 
\draw[pattern = north east lines] (8,0) circle (20 pt); 
\draw[pattern = north east lines] (10,0) circle (20 pt); 
\draw[pattern = north east lines] (12,0) circle (20 pt); 
\draw (1.2,1.6) circle (20 pt); 
%\draw (3,1.6) circle (20 pt); 
\draw (5.1,1.5) circle (20 pt); 
\draw (7.1,1.6) circle (20 pt); 
\draw (9.3,1.2) circle (20 pt); 
\draw (11.3,1.4) circle (20 pt); 
\draw[pattern = north east lines] (0,3.2) circle (20 pt); 
\draw (2.4,3.1) circle (20 pt); 
\draw (4.3,3.3) circle (20 pt); 
\draw (6.35,3.2) circle (20 pt); 
\draw (7.9,3.2) circle (20 pt); 
\draw (10.45,3.6) circle (20 pt); 
\draw (9.2,2.65) circle (20 pt); 
\draw[pattern = north east lines] (12,3.2) circle (20 pt); 
\draw (1,4.7) circle (20 pt); 
\draw (3.5,4.8) circle (20 pt); 
\draw (5.9,4.9) circle (20 pt); 
\draw (7.5,4.9) circle (20 pt); 
\draw (9.4,5) circle (20 pt); 
\draw (11.3,4.9) circle (20 pt); 
\draw[pattern = north east lines] (0,6.4) circle (20 pt); 
\draw (2.4,6.4) circle (20 pt); 
\draw (4.6,6.3) circle (20 pt); 
\draw[pattern = dots] (7.0,6.4) circle (20 pt); 
\draw (8.7,6.5) circle (20 pt); 
\draw (10.3,6.6) circle (20 pt); 
\draw[pattern = north east lines] (12,6.4) circle (20 pt); 
\draw (1.3,8.1) circle (20 pt); 
\draw (3.4,8) circle (20 pt); 
\draw (5.6,8.2) circle (20 pt); 
\draw (7.9,7.9) circle (20 pt); 
%\draw (9,8) circle (20 pt); 
\draw (10.9,8.2) circle (20 pt); 
\draw[pattern = north east lines] (0,9.6) circle (20 pt); 
%\draw (2,9.6) circle (20 pt); 
\draw (4.2,9.7) circle (20 pt); 
\draw (6.5,9.5) circle (20 pt); 
\draw (8.2,9.6) circle (20 pt); 
\draw (10.1,9.6) circle (20 pt); 
\draw[pattern = north east lines] (12,9.6) circle (20 pt); 
\draw (0.82,11.4) circle (20 pt); 
\draw (3.15,11.3) circle (20 pt); 
\draw (5.25,11.2) circle (20 pt); 
\draw (7.15,11.4) circle (20 pt); 
\draw (9.35,11.2) circle (20 pt); 
\draw (10.95,11.1) circle (20 pt); 
\draw[pattern = north east lines] (0,12.8) circle (20 pt); 
\draw[pattern = north east lines] (2,12.8) circle (20 pt); 
\draw[pattern = north east lines] (4,12.8) circle (20 pt); 
\draw[pattern = north east lines] (6,12.8) circle (20 pt); 
\draw[pattern = north east lines] (8,12.8) circle (20 pt); 
\draw[pattern = north east lines] (10,12.8) circle (20 pt); 
\draw[pattern = north east lines] (12,12.8) circle (20 pt); 
%\draw[color=red] (2,1.65) circle (11 pt); 
\end{tikzpicture}
\end{center}
\caption{Illustration of the transformation $\Tn$. The large square is $\La_n$, 
the small square is of size $\sqrt n$. The disks forming the 
boundary configuration are shaded. A particle near the center is dotted. 
The left hand side shows a configuration $X$, 
the right hand side shows the corresponding configuration $\Tn(X)$: 
Every particle $x \in X$ is moved some distance $\tnx(x)$ to the right.
This distance may depend on the initial position $x$ as well as on the part 
of the configuration $X$ surrounding $x$.  
} 
\label{figTn}
\end{figure}

\newpage 

In the above theorem $\Tn$ should be thought of a transformation
shifting every particle $x$ of a given configuration $X$ by the amount $\tnx(x)$ 
in direction $\einh$, see Figure \ref{figTn}. The middle region $\La_{\sqrt n}$ 
is shifted to the right by $\de \ep \sqrt{\log n}$. The particle density in the left 
part of $\La_n$ is decreased and the particle density in the right part of $\La_n$ 
is increased. 
The above properties of the transformation can be interpreted as follows: 
By (1) the transformation does not affect particles outside of $\Lan$, i.e. 
particles of the boundary configuration.  
By (2) particles near the origin are shifted by an amount of order $\sqrt{\log n}$
provided the configuration is good. 
By (3) particles that are close to each other are shifted by almost the same amount. 
In particular, if particles locally form a lattice, 
this lattice structure is almost preserved by the transformation.  
By (4) good configurations are likely. 
By (5) the transformation is bijective and compatible with the hard core condition.  
By (6) we have control over probabilities when applying the transformation. 
By (7) the price to pay when applying the transformation in terms of 
change of probabilities does not depend on $n$.
The reason why we consider the transformation in some direction 
along with the corresponding transformation in the opposite direction is that 
we are not able to obtain the corresponding estimate for $\log \ph$ alone
(similar to Mermin-Wagner-type arguments). 
We note that it is necessary to introduce a set of good configurations, 
because otherwise the above properties are in conflict: For a configuration 
with a dense packing of particles to the right of $\La_{\sqrt n}$ 
properties (1) and (6) imply that there is not enough room for particles 
in $\La_{\sqrt n}$ to be moved by the target amount. Configurations like 
that should be thought of as bad. 

\bigskip

Since the hard disk model does not have an a priori lattice structure, it is not 
clear how to refer to a specific particle and make statements about the variance 
of its position or the displacement from its ideal lattice position. 
We would like to argue that in a case like this a theorem as the above 
is a substitute for these statements. Indeed, Theorem \ref{thmtransformation}
guarantees that it is possible to displace all particles in some region 
by an amount of order  $\sqrt{\log n}$ while preserving local structures, 
and this can be done without affecting probabilities significantly. 
The following corollary is meant to motivate this interpretation. 
It shows that if there is an a posterioi lattice structure by which we may 
identify some particle $\xi \in X$, whose ideal lattice position is close to the origin, 
then the mean square deviation of its position 
from the ideal lattice position is at least of order $\log n$. The proof of this 
result it will also explain the role of property (7). 
\begin{kor} \label{corfluctuations}
For  $z > 0$, $\ep := \min\{\frac 1 {48z},\frac 1 4\}$, $\de \in (0,\frac 1 {30}]$  and $n \ge 200$
let $\Tn, \iTn$ be transformations as in Theorem \ref{thmtransformation}. 
Let $\xi: \XX \to \R^2 \cup \{\ho\}$ be a rule for picking a particle from a 
configuration, i.e. $\xi(X) \in X \cup \{\circ\}$ for all $X \in \XX$, 
such that $\mu_n^z(\xi \neq \circ|\bX) > \frac 1 2$,  and such that 
$\xi$ is compatible with $\Tn$ in that $\xi(\Tn(X)) = \xi(X) + \tnx(\xi(X))\einh$ 
for all $X \in \XX$ and similarly for $\iTn$. 
Then for every $\bxi \in  \La_{\sqrt n/2}$ and  $Y \in \XX$ we have 
$$
\mu_n^z(|\xi- \bxi| \ge  \frac{\de \ep}2 \sqrt{\log n}|Y) \ge \frac 1 8  \; 
\text{ and thus } \;
\mu^n_z(|\xi -\bxi|^21_{\{\xi \neq \circ\}}|\bX) \ge  \frac{\de^2 \ep^2}{32} \log n.
$$
\end{kor}

\newpage 

Here $|.|$ denotes the maximum norm. 
The value $\circ$ of $\xi$ corresponds to a situation where particle $\xi$ 
is absent in a configuration or can't be chosen unambiguously. 
While the particle labelling  mechanism $\xi$ in the above corollary is quite general, 
we would like to think of $\bX$ forming a perfect triagonal lattice with a particle 
density corresponding to the given value of $z$, we would like to think of $\mu^z_n$ 
as producing particle configurations that have a lattice structure similar to 
the one of $\bX$ in that most particles can be labelled by corresponding lattice sites 
in a consistent way. 
$\xi$ should pick the particle corresponding to the lattice site $\bxi$. 
We note that Figure~\ref{figTn} depicts a situation like that. Here $\bxi$ is the 
lattice site at the center of the lattice formed by the boundary configuration. 
The corresponding disk $\xi$ is marked as dotted in both configurations. 
We stress that it is not clear to which extent the hard core model 
exhibits a lattice structure. The corollary is merely meant to illustrate 
that whenever we are able to refer to a particle, 
the theorem indeed implies a lower bound on the fluctuation of its positions. 

\bigskip 

We note that we have restricted ourselves to the case of pure hard core repulsion only 
to simplify the exposition of the proof. In the following we describe various possible 
generalizations. For details on the corresponding setting and definitions 
we refer to \cite{R} and \cite{R2}. Indeed, the proof of the generalizations
consists of a combination of the ideas used in this paper and the technical machinery  of \cite{R} and \cite{R2}. Details will be provided in a forthcoming paper. 
Theorem \ref{thmtransformation} still holds for the following generalizations: 
\begin{itemize}
\item $U$ may be nonvanishing outside of the hard core. 
$U$ still has to be symmetric and translation invariant, and
outside of the hard core $U$ has to be smooth with a certain 
integrability condition on the second derivative  ($\psi$-dominated derivatives 
for a decay function $\psi$ in the sense of \cite{R}). 
\item 
$U$ may have a hard core of a different shape or no hard core at all. 
Instead it may have a singularity or be bounded.
In addition to the conditions above, here $U$ needs to admit a Ruelle bound 
(in the sense of \cite{R}), e.g. $U$ nonnegative or $U$ super-stable and lower 
regular. 
\item $U$ may not be smooth outside of the hard core/singularity, 
but may have discontinuities, and the way 
$U$ behaves near the hard core/singularity is not relevant. 
What we need here is a symmetric, translation-invariant potential $U$ 
that admits a Ruelle bound and is smoothly approximable in the sense of 
Definition 1 of \cite{R}. 
\item 
We may have different types of particles (e.g. hard disks with different radii), 
or particles with internal degrees of freedoms (e.g. hard squares or hard rods 
with random orientation). For a formulation of the conditions on $U$ 
in this case, see Definition 1 of \cite{R2}. 
\end{itemize}

\newpage

\section{Fluctuations of particle positions: Corollary \ref{corfluctuations}}
\label{secproofcor}

Here we deduce Corollary \ref{corfluctuations} from Theorem \ref{thmtransformation}. 
Let  $z > 0$, $\ep := \min\{\frac 1 {48z},\frac 1 4\}$, $\de \in (0,\frac 1 {30}]$ and $n \ge 200$. 
Let $\Tn, \iTn$ transformations as in Theorem \ref{thmtransformation}.
Let $\xi: \XX \to \R^2 \cup \{\ho\}$ be a rule for picking particles from configurations
with the given properties, let $\bxi \in \La_{\sqrt n/2}$ and $Y \in \XX$.  
Let 
\begin{align*}
&E_n := \{|\xi- \bxi| \ge  \frac{\de \ep}2 \sqrt{\log n}\}, 
 \quad D_n := \{|\xi - \bxi| < \frac{\de \ep}2 \sqrt{\log n}\} \\
&\quad \text{ and } D_n^\pm := \{|\xi- \bxi \pm \de \ep \sqrt{\log n} \einh| < \frac{\de \ep}2 \sqrt{\log n}\}.
\end{align*}
All four events tacitly imply $\xi \in \R$, i.e. $\xi \neq \circ$.
We note that for all $X \in D_n \cap G_n$ we have $\xi(X)  \in \La_{\sqrt n}$
since $\bxi \in \La_{\sqrt n/2}$. Thus property (3) of the transformation 
implies $\tnx(\xi(X)) = \de \ep \sqrt{\log n}$. 
By the compatibility property of $\xi$ this gives 
$\xi(\Tn(X)) = \xi(X) + \de \ep \sqrt{\log n} \einh$, 
i.e. $\Tn(X) \in D_n^-$. 
Thus we have shown that $\Tn(D_n \cap G_n) \subset D_n^-$.
This implies 
\begin{align*}
&\mu_n^z(D_n^-|Y) \ge  \mu_n^z(\Tn(D_n \cap G_n)|Y)
= \int \mu_n^z(dX|Y) 1_{\Tn(D_n \cap G_n)}(X)\\ 
&= \int \mu_n^z(dX|Y) 1_{\Tn(D_n \cap G_n)}(\Tn(X))\ph(X)
= \int \mu_n^z(dX|Y) 1_{D_n \cap G_n}(X)\ph(X)
\end{align*}
using property (4) of the transformation. Combining this with the corresponding 
estimate for $\iTn$ and the inequality 
$$
\ph(X) + \bph(X) \ge 2 \sqrt{\ph(X)\bph(X)} \ge 1_{\sqrt{\ph\bph} \ge \frac 1 2}(X)
$$
we obtain
\begin{align*}
&\mu_n^z(E_n|Y) \ge  \mu_n^z(D_n^- \cup D_n^+|Y) 
\ge \int \mu_n^z(dX|Y) 1_{D_n \cap G_n}(X)(\ph(X)+\bph(X))\\
&\ge \mu_n^z( D_n \cap G_n \cap \{\sqrt{\ph\bph} \ge \frac 1 2\}|Y) \\
&\ge \mu_n^z(D_n|Y) - \mu_n^z(G_n^c|Y) - \mu_n^z(\sqrt{\ph\bph} \le \frac 1 2|Y).  
\end{align*}
Property (4) of the transformation implies 
that $\mu_n^z(G_n^c|Y) \le \frac 1 8$, property (5) implies that 
\begin{align*}
&\mu_n^z(\sqrt{\ph\bph} \le \frac 1 2|Y) \le \mu_n^z(|\log(\ph\bph)| \ge \log 4|Y) 
\le \frac{120 \de^2}{\log 4} \le \frac 1 8
\end{align*}
and by the given properties of $\xi$ we have 
$$
\mu_n^z(E_n|Y) + \mu_n^z(D_n|Y) = \mu_n^z(\xi \neq \circ|Y) \ge \frac 1 2, 
$$
so the above implies 
$$
\mu_n^z(E_n|Y) \ge \frac 1 2 - \mu_n^z(E_n|Y) - \frac 1 8 - \frac 1 8 \quad \gdw  \quad 
\mu_n^z(E_n|Y) \ge \frac 1 8. 
$$
This gives the first estimate and the second estimate is a direct consequence.

\section{Transformation: Proof of Theorem \ref{thmtransformation}} \label{secproofthm}

\begin{figure}[!b]
\begin{center}
\quad \begin{tikzpicture}[scale = 1]
\draw[->] (-5,0) -- (4,0) node[right]{$\R^2$ \hspace*{0.6 cm}}; 
\draw[|-|] (-3,-1)-- (3,-1) node[right]{$\Lan$};
\draw[|-|] (-1,-1.5)-- (1,-1.5) node[right]{$\La_{n^{2/3}}$};
\draw(-4.5,0) circle (0.3);
\draw[->](-4.5,0.5)--(-4.45,0.5); 
\draw(3.2,0) circle (0.3);
\draw[->](3.2,0.5)--(3.25,0.5); 
\draw(- 2.5,0) circle (0.3);
\draw[->](-2.55,0.5)-- (-2.45,0.5); 
\draw(-0.2,0) circle (0.3);
\draw[->](-0.4,0.5)--(0,0.5); 
\draw(1.4,0) circle (0.3);
\draw[->](1.3,0.5)--(1.5,0.5); 
\draw(2.3,0) circle (0.3);
\draw[->](2.25,0.5)--(2.35,0.5); 
\end{tikzpicture}

\quad\begin{tikzpicture}[scale = 1]
\draw[->] (-5,0) -- (4,0) node[right]{$x \in \R^2$};
\draw[->] (0,0) -- (0,1.3) node[above]{$t_n^0(x)$};
\draw (-3,0) -- (-1,1)--(1,1)--(3,0);
%\draw (-0.1,1) -- (0.1,1) node[above left]{$\tau$};
\draw (-1,0.1) -- (-1,-0.1);  
\draw (-3,0.1) -- (-3,-0.1);  
\draw (1,0.1) -- (1,-0.1) node[below]{$n^{2/3}$};  
\draw (3,0.1) -- (3,-0.1) node[below]{$n$};  
\draw[fill](-4.5,0) circle (0.05);
\draw[fill](3.2,0) circle (0.05);
\draw[fill](- 2.5,0) circle (0.05);
\draw[fill](-0.2,0) circle (0.05);
\draw[fill](1.4,0) circle (0.05);
\draw[fill](2.3,0) circle (0.05);
\end{tikzpicture}
\end{center}
\caption{Illustration of the shift profile in a 1D situation. 
The top part shows a configuration $X$ of disks. The dimensions 
of the boxes $\La_n$ and $\La_{n^{2/3}}$ are indicated. For every particle 
$x \in X$ the translation distance $\etn^0(x)$ is indicated by the arrow
above $x$. The lower part shows $\etn^0$ in terms of a shift profile. 
Here the particle positions are indicated by points. The flat part of $\etn^0(x)$  
near the origin has height $\de \ep \sqrt{\log n}$. For the sake of the 
illustration the interpolation \eqref{deftn} has been replaced by a linear 
interpolation.}  
\label{figtau}
\end{figure}
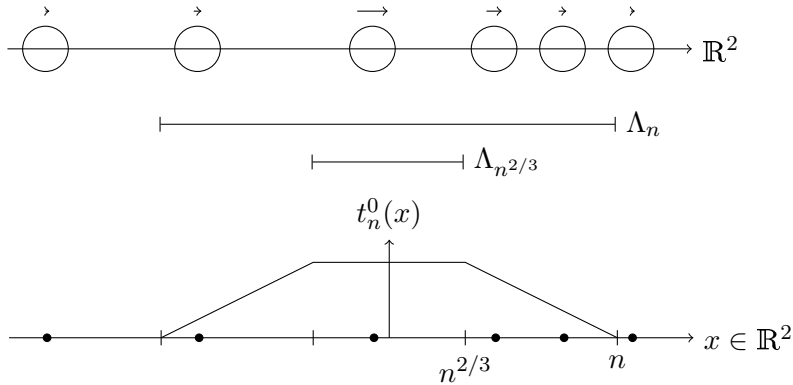

Let  $z > 0$, $\ep := \min\{\frac 1 {48z},\frac 1 4\}$ and $\de \in (0,\frac 1 2]$. 
These parameters will be fixed throughout this proof 
and dependencies on these parameters will be suppressed. 
Let  $n \ge 200$. We will construct a corresponding transformation $\Tn$ 
and a set of good configurations $\Gn$ satisfying properties (1) - (7) from 
Theorem~\ref{thmtransformation}. 
%We start by defining a certain interpolation profile of the translation distance. 
Let $\tn: \R \to [0,\infty)$ be defined by 
\begin{equation}\label{deftn}
\tn(s) \, := \,\left\{  
\begin{aligned}
& \de \ep \sqrt{\log n} && \text{ for } s \le n^{2/3}\\
&\frac {3 \de \ep} {\sqrt{\log n}} (\log n - \log s) && \text{ for } n^{2/3} \le s \le n\\
& 0 &&\text{ for } s \ge n. 
\end{aligned}\right.
\end{equation}
%We note that $\tn$ is continuous and decreasing. 
We take $\etn^0 \, := \, \tn(|.|)$ to be a first approximation 
of the translation distance function $\tnx$. Here $|.|$ denotes the maximum norm on $\R^2$.
$\etn^0: \R^2 \to [0,\infty)$ specifies how far a particle at a given position 
would like to be shifted in direction $\einh$ and we call such a function a shift profile, see Figure \ref{figtau}. 
According to the shift profile $\etn^0$ every particle is shifted by an amount, 
which only depends on its position. It can be seen that the corresponding 
transformation satisfies all properties but (3): Hard cores are not respected
since particles that are almost at hard core distance may be shifted by different 
amounts, which may cause collisions of hard cores. To include property (3), 
the idea is that every particle should have an effect on all other particles 
within or close to hard core distance, slowing down the shift of these particles, 
and thus preventing collisions. This slow down of the shift can be achieved 
by modifying the shift profile accordingly. The slow down caused by 
a particle at position $p \in \R^2$ which is known to be shifted by the amount $t \ge 0$
can be implemented by a function  $\mxt: \R^2 \to [0,\infty) \cup \{\infty\}$ of the following type: 
$$
\mxt(x) \, := \, \left\{
\begin{aligned}
&t && \text{ for } | x-p |_2 \le 1 \\
& t +  \frac \hxt \ep  (| x-p |_2-1)  && \text{ for } 1 \le | x-p |_2  \le 1 + \ep \\ 
&\infty && \text{ for } | x-p |_2  > 1 + \ep 
\end{aligned} \right. 
$$ 
Here 
$$
 \hxt \, :=  \, |\tn(|p| - 1-\ep) - t|.
$$
is the maximum possible shift amount of a particle at distance $1 + \ep$ from $p$ 
as proposed by $\tn$ in comparison to the shift amount of the particle at $p$, 
see Figure~\ref{figm}. 
(One should think of the case $\tn(|p|) \ge t$.) 
The parameter $\ep = \min\{\frac 1 {48 z}, \frac 1 4\}$
regulates the range in which the slow down is felt. 
Functions of the above type will be used to locally modify shift 
profiles causing slow downs wherever needed. However, these slow downs will prevent collisions
only if the slope of these functions can be controlled. So we set 
\begin{equation} \label{altdefm}
\mxt(x) \, := \, t \text{ for all } x \in \R \quad \text{ if } \hxt  > \de \ep.
\end{equation}
For good configurations this proviso will turn out not be necessary. 
The details of the above construction are chosen to guarantee the 
following property: 
\begin{lem} \label{lepropm} Let $p \in \R^2$ and $t \ge 0$. 
The pointwise minimum of $\etn^0$ and $\mxt$ 
is a Lipschitz-continuous function with Lipschitz-constant $\de$. 
\end{lem}
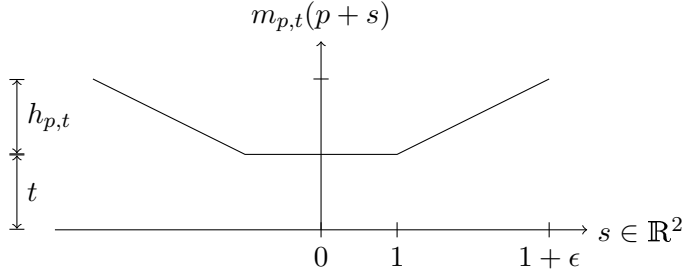
\begin{figure}[!htb]
\begin{center}
\begin{tikzpicture}[scale = 1]
\draw[->] (-3.5,0) -- (3.5,0)  node[right]{$s \in \R^2$};
\draw[->] (0,0) -- (0,2.5) node[above]{$\mxt(p+s)$}; 
\draw (-3,2) -- (-1,1)--(1,1) --(3,2);
\draw (0,0.1) -- (0,-0.1) node[below]{$0$};  
\draw (1,0.1) -- (1,-0.1) node[below]{$1$};  
\draw (3,0.1) -- (3,-0.1) node[below]{$1 + \ep$};  
\draw(-0.1,2) -- (0.1,2);
\draw[|<->|] (-4,1) --  node[right]{$\hxt$}(-4,2);
\draw[|<->|] (-4,0) --  node[right]{$t$}(-4,1);
\end{tikzpicture}
\end{center}
\caption{1D illustration of the function $\mxt$ in case of $\hxt \le \de \ep$.}
\label{figm}
\end{figure}

Now we are ready to define the transformation $\Tn: \X \to \X$. Let $X \in \X$ 
and $m(X)$ be the number of particles of $X_{\La_n}$. 
We define $\Tn(X)$ by recursively constructing the following objects for 
$1 \le k \le m(X)$, see Figure \ref{figtnx}:  
\begin{itemize}
\item An enumeration $\pnx^k$ of the particles of $X_{\La_n}$. 
\item Shift amounts $\taunx^k \in [0,\infty)$ for the particles $\pnx^k$.
\item Shift profiles $\tnx^k: \R^2 \to [0,\infty)$ that take into account the slow down 
due to an increasing number of particles. 
\end{itemize}
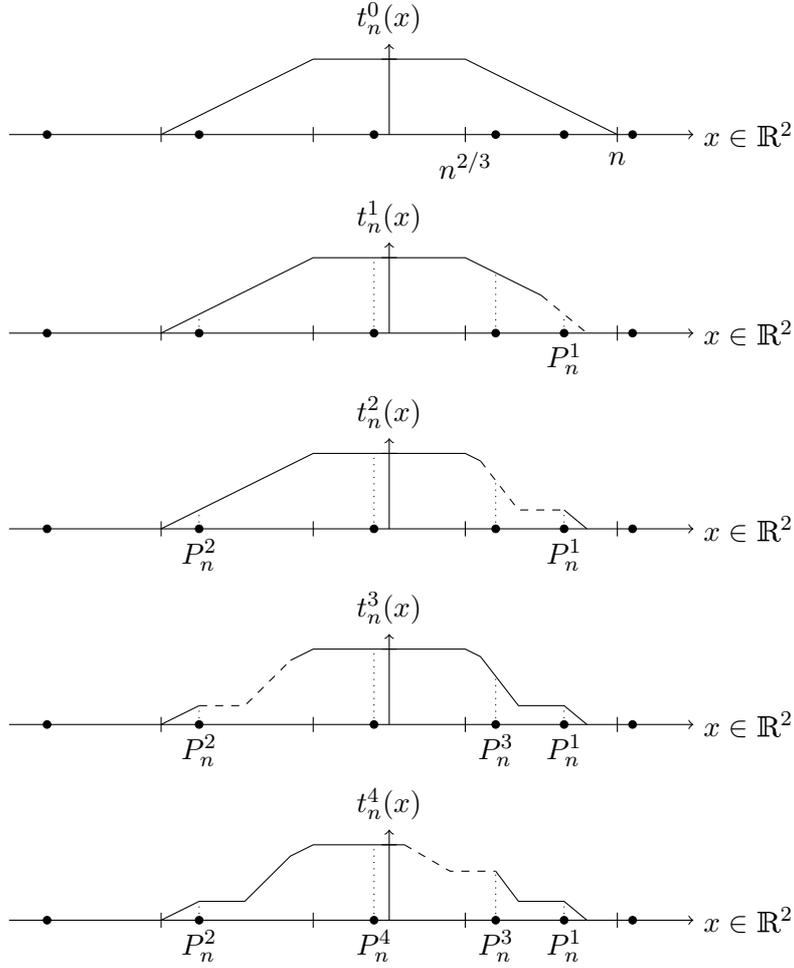
\begin{figure}[!htb]
\begin{center}
\quad\begin{tikzpicture}[scale = 1]
\draw[->] (-5,0) -- (4,0) node[right]{$x \in \R^2$};
\draw[->] (0,0) -- (0,1.2) node[above]{$\etn^0(x)$};
\draw (-3,0) -- (-1,1)--(1,1)--(3,0);
\draw (-0.1,1) -- (0.1,1); % node[above left]{$ \tau$};
%\draw (-0.5,0.1) -- (-0.5,-0.1);  
\draw (-1,0.1) -- (-1,-0.1);  
\draw (-3,0.1) -- (-3,-0.1);  
%\draw (0.5,0.1) -- (0.5,-0.1) node[below]{$ n'$};  
\draw (1,0.1) -- (1,-0.1) node[below]{$n^{2/3}$};  
\draw (3,0.1) -- (3,-0.1) node[below]{$ n$};  
\draw[fill](-4.5,0) circle (0.05);
\draw[fill](3.2,0) circle (0.05);
\draw[fill](- 2.5,0) circle (0.05);
\draw[fill](-0.2,0) circle (0.05);
\draw[fill](1.4,0) circle (0.05);
\draw[fill](2.3,0) circle (0.05);
\end{tikzpicture}

\quad 
\begin{tikzpicture}[scale = 1]
\draw[->] (-5,0) -- (4,0) node[right]{$x \in \R^2$};
\draw[->] (0,0) -- (0,1.2) node[above]{$\etn^1(x)$};
\draw (-3,0) -- (-1,1)--(1,1)--(2,0.5); 
\draw[dashed] (2,0.5) -- (2.6,0);
\draw (-0.1,1) -- (0.1,1);
%\draw (-0.5,0.1) -- (-0.5,-0.1);  
\draw (-1,0.1) -- (-1,-0.1);  
\draw (-3,0.1) -- (-3,-0.1);  
%\draw (0.5,0.1) -- (0.5,-0.1);  
\draw (1,0.1) -- (1,-0.1);  
\draw (3,0.1) -- (3,-0.1);  
\draw[fill](-4.5,0) circle (0.05);
\draw[fill](3.2,0) circle (0.05);
\draw[fill](- 2.5,0) circle (0.05);
\draw[dotted](-2.5,0) -- (-2.5,0.25);
\draw[fill](-0.2,0) circle (0.05);
\draw[dotted](-0.2,0) -- (-0.2,1);
\draw[fill](1.4,0) circle (0.05);
\draw[dotted](1.4,0) -- (1.4,0.8);
\draw[fill](2.3,0) circle (0.05) node[below]{$ \epn^1$};
\draw[dotted](2.3,0) -- (2.3,0.25);
\end{tikzpicture}

\quad 
\begin{tikzpicture}[scale = 1]
\draw[->] (-5,0) -- (4,0) node[right]{$x \in  \R^2$};
\draw[->] (0,0) -- (0,1.2) node[above]{$\etn^2(x)$};
\draw (-3,0) -- (-1,1)--(1,1)--(1.2,0.9); 
\draw[dashed] (1.2,0.9) -- (1.7,0.25) -- (2.3,0.25);
\draw (2.3,0.25) -- (2.6,0);
\draw (-0.1,1) -- (0.1,1);
%\draw (-0.5,0.1) -- (-0.5,-0.1);  
\draw (-1,0.1) -- (-1,-0.1);  
\draw (-3,0.1) -- (-3,-0.1);  
%\draw (0.5,0.1) -- (0.5,-0.1);  
\draw (1,0.1) -- (1,-0.1);  
\draw (3,0.1) -- (3,-0.1);  
\draw[fill](-4.5,0) circle (0.05);
\draw[fill](3.2,0) circle (0.05);
\draw[fill](- 2.5,0) circle (0.05)  node[below]{$\epn^2$};
\draw[dotted](-2.5,0) -- (-2.5,0.25);
\draw[fill](-0.2,0) circle (0.05);
\draw[dotted](-0.2,0) -- (-0.2,1);
\draw[fill](1.4,0) circle (0.05);
\draw[dotted](1.4,0) -- (1.4,0.65);
\draw[fill](2.3,0) circle (0.05) node[below]{$ \epn^1$};
\draw[dotted](2.3,0) -- (2.3,0.25);
\end{tikzpicture}

\quad
\begin{tikzpicture}[scale = 1]
\draw[->] (-5,0) -- (4,0) node[right]{$x \in \R^2$};
\draw[->] (0,0) -- (0,1.2) node[above]{$\etn^3(x)$};
\draw (-1.3,0.85) -- (-1,1)--(1,1)--(1.2,0.9) -- (1.7,0.25) -- (2.3,0.25)-- (2.6,0);
\draw (-3,0)--(-2.5,0.25);
\draw[dashed] (-2.5,0.25)--(-1.9,0.25) -- (-1.3,0.85);
\draw (-0.1,1) -- (0.1,1);
%\draw (-0.5,0.1) -- (-0.5,-0.1);  
\draw (-1,0.1) -- (-1,-0.1);  
\draw (-3,0.1) -- (-3,-0.1);  
%\draw (0.5,0.1) -- (0.5,-0.1);  
\draw (1,0.1) -- (1,-0.1);  
\draw (3,0.1) -- (3,-0.1);  
\draw[fill](-4.5,0) circle (0.05);
\draw[fill](3.2,0) circle (0.05);
\draw[fill](- 2.5,0) circle (0.05)  node[below]{$ \epn^2$};
\draw[dotted](-2.5,0) -- (-2.5,0.25);
\draw[fill](-0.2,0) circle (0.05);
\draw[dotted](-0.2,0) -- (-0.2,1);
\draw[fill](1.4,0) circle (0.05)  node[below]{$ \epn^3$};
\draw[dotted](1.4,0) -- (1.4,0.65);
\draw[fill](2.3,0) circle (0.05) node[below]{$ \epn^1$};
\draw[dotted](2.3,0) -- (2.3,0.25);
\end{tikzpicture}

\quad 
\begin{tikzpicture}[scale = 1]
\draw[->] (-5,0) -- (4,0) node[right]{$x \in \R^2$};
\draw[->] (0,0) -- (0,1.2) node[above]{$ \etn^4(x)$};
\draw (-3,0)--(-2.5,0.25)--(-1.9,0.25) --  (-1.3,0.85) -- (-1,1)--
(0.2,1);
\draw (1.4,0.65) -- (1.7,0.25) -- (2.3,0.25)-- (2.6,0);
\draw[dashed] (0.2,1) -- (0.8,0.65) -- (1.4,0.65);
\draw (-0.1,1) -- (0.1,1);
%\draw (-0.5,0.1) -- (-0.5,-0.1);  
\draw (-1,0.1) -- (-1,-0.1);  
\draw (-3,0.1) -- (-3,-0.1);  
%\draw (0.5,0.1) -- (0.5,-0.1);  
\draw (1,0.1) -- (1,-0.1);  
\draw (3,0.1) -- (3,-0.1);  
\draw[fill](-4.5,0) circle (0.05);
\draw[fill](3.2,0) circle (0.05);
\draw[fill](- 2.5,0) circle (0.05)  node[below]{$ \epn^2$};
\draw[dotted](-2.5,0) -- (-2.5,0.25);
\draw[fill](-0.2,0) circle (0.05)  node[below]{$ \epn^4$};
\draw[dotted](-0.2,0) -- (-0.2,1);
\draw[fill](1.4,0) circle (0.05)  node[below]{$ \epn^3$};
\draw[dotted](1.4,0) -- (1.4,0.65);
\draw[fill](2.3,0) circle (0.05) node[below]{$ \epn^1$};
\draw[dotted](2.3,0) -- (2.3,0.25);
\end{tikzpicture}
\end{center}
\caption{1D illustration of the recursive construction. The positions of the 
particles are indicated by dots.}
\label{figtnx}
\end{figure}
In the above notations we will omit the dependence on $X$ whenever it is 
clear which configuration $X$ is considered. For fixed $X \in \X$ 
the $k$-th step  $(1 \le k \le m)$ of our recursive construction is the following:  
\begin{itemize}
\item We set the shift profile $\etn^k$ to be the minimum of $\etn^{k-1}$ and the slow down 
$m_{\epn^{k-1},\etaun^{k-1}}$. Here $\etn^0 = \tn(|.|)$ as above and 
$m_{\epn^{0},\etaun^{0}}$ is the minimum of all slow downs $m_{x,0}$ where $x \in X_{\La_n^c}$. 
\item  Let $\epn^k$ be the point of $X_{\Lan} \weg \{\epn^1, \ldots, \epn^{k-1}\}$ 
at which the minimum of $\etn^{k}$ is attained. 
If there is more than one such point then take the smallest point 
with respect to the lexicographic order for the sake of definiteness. 
\item 
Let  $\etaun^k := \etn^{k}(\epn^k)$  be the corresponding minimal value of 
$\etn^k$.
\end{itemize}
We also set $\etaun^0 := 0$ and $|\epn^0| := n$ and think of $\epn^0$ as an arbitrary 
point of $X_{\Lan^c}$. 
The transformation $\Tn: \X \to \X$ is now defined by 
\begin{align*}
&\Tn(X) \, := \, \{x + \tnx(x) \einh : x \in X \}, \quad \text{ where }\\
&\tnx(x) = 0 \text{ for } x \in X_{\La_n^c} \quad \text{ and } \quad 
\tnx(\pnx^k) = \taunx^k \text{ for } 1 \le k \le m(X). 
\end{align*}
For a formulation of suitable properties of the construction we need some more 
notation. In the above construction 
$\etn^k(\epn^k)$ is the minimum of $\etn^0(\epn^k)$ and the functions 
$m_{\epn^l,\etaun^l}(\epn^k)$ ($0 \le l < k$). For $0 \le l < k$ we write 
$$
\epn^k \to \epn^l \quad \text{ if } \etn^k(\epn^k) = m_{\epn^l,\etaun^l}(\epn^k),
$$
i.e. loosely speaking if the shift of $\epn^k$ is determined by the point $\epn^l$, 
and $\epn^k \to \emptyset$ if $\etn^k(\epn^k) = \etn^0(\epn^k)$, i.e. if 
the shift of $\epn^k$ does not depend on other points.  Let 
$$
\anx(\epn^k) := \epn^i, \text{ where } i := \min\{ 0 \le l < k: \epn^k \to ... \to \epn^l\}, 
$$
denote the first point of the construction that has an indirect influence on the shift of $\epn^k$
(via a sequence of other points). If the set is empty we set $\anx(\epn^k) := \epn^k$. 
We also define 
$$
m'(X) := \min\{0 \le l \le m: h_{\epn^l,\etaun^l} >  \ep \de \}
$$
to be the first time in the construction that the proviso \eqref{altdefm} in the definition 
of $m_{p,t}$ has to be used. If the set is empty we set $m'(X) := m+1$. 

\bigskip 

In the following sequence of lemmas we show that the above construction gives a transformation 
$\Tn$ with the desired properties. We start with properties that are almost immediate 
from the construction. 
\begin{lem}\label{leeasyprop} Let $X \in \X$ be a configuration. We have the
monotonicity  properties: 
\begin{align}   
&0 = \etaun^0 \le \etaun^1 \le \ldots \le \etaun^{m}, \label{mono1}\\
&\etn^0 \ge \etn^1 \ge ... \ge \etn^m \ge 0, \label{mono2}
\end{align}
and consequently 
\begin{equation} \label{tnxalt}
\forall 0 \le k \le l \le m: \etn(\epn^k) = \etaun^k = \etn^l(\epn^k).
\end{equation}
Analysing which point has an influence on the shift of which other points we get 
\begin{align} 
&\forall x \in X: 
\tn(|\anx(x)|) \le \etn(x) \le \tn(|x|)  \label{tauab}\\
%&\forall k > m': \etaun^k = \etaun^{m'} \text{ and } \etn^k = \etn^{m'+1}, \label{proviso1}\\
&\forall 0 \le l< k \le m': \epn^k \to \epn^l \; \Rightarrow \; |\epn^k - \epn^l|_2 \le 1+\ep. \label{proviso2}   
\end{align}
We have the following continuity property:  
\begin{equation}
\text{Each $\etn^k$ ist Lipschitz-continuous with Lipschitz constant $\de$.}  \label{Lcont}
\end{equation}
For all particles $x_1,x_2 \in X$ possible hard cores are respected, i.e.  
\begin{align}
&|x_1-x_2|_2 \le 1  \; \Rightarrow \; \etn(x_1) \,= \,\etn(x_2), \label{hc1}\\
&|x_1-x_2|_2 > 1 \; \Rightarrow \; |(x_1+ \etn(x_1)\einh)-(x_2+ \etn(x_2)\einh)|_2 > 1. \label{hc2} 
\end{align}
\end{lem}
We note that property (1) of the transformation is satisfied by construction. 
Property (3) is a consequence of the Lipschitz-continuity
\eqref{Lcont} for $\etn^m$ and $\etn^m = \etn$ from \eqref{tnxalt}. 
The first part of property (5) is shown in the following lemma: 
\begin{lem} \label{lebij}
The transformation $\Tn: \X \to \X$ is bijective.
\end{lem}
Since $\Tn(\XX) \subset \XX$ and $\Tn((\XX)^c) \subset (\XX)^c$ by \eqref{hc1} and \eqref{hc2}, 
we obtain the second part of property (5). For property (6) we set 
\begin{equation*} 
\ph(X) \, := \, 
\prod_{k = 1}^{m(X)} \big|1 +  \partial_{\einh} \tnx^{k} (\pnx^k)\big| 
\end{equation*}
for $X \in \X$. While proving that this is indeed the density we are looking for, 
we will also show that this definition makes sense 
$\mu_n^z (\,. \, |\bX)$-a.s., in that the considered partial derivatives exist. 
\begin{lem} \label{ledensphc}
For every  $\bX \in \XX$ and every measurable function $f \ge 0$ 
\begin{equation*} 
\int d\mu_n^z(dX|\bX) \, f(\Tn(X))\ph(X) \,
= \, \int d\mu_n^z (dX|\bX) \, f(X).
\end{equation*}
\end{lem}
This completes the proof of property (6). For property (7) we consider 
the transformation  $\iTn$, shifting particles by the same amount in the 
opposite direction. By symmetry it has properties analogous to those 
of $\Tn$. We note that $\iTn$ is not the inverse of $\Tn$. We now show 
property (7). 
\begin{lem} \label{ledensest}
For every $\bX \in \XX$ we have  
\begin{equation*}
\mu_n^z( |\log(\ph \bph)||\bX) \le 120 \de^2. 
\end{equation*}
\end{lem}
The remaining properties (2) and (4) concern good configurations. 
In light of \eqref{tauab}, property (2) follows provided we have sufficient control 
over $\anx(x)$ for all $x$. To achieve this control, in light of \eqref{proviso2} 
we  compare the relation $\to$ to continuum percolation 
of disks. For $x,x' \in X_{\Lan}$ we set 
\begin{align*}
&x \sim x' \quad :\gdw \quad |x-x'|_2 \le 1+\ep \quad \text{ and }\\
&\rnx(x) : = \max\{|x'|: x \sim ... \sim x'\}.
\end{align*}
$\rnx(x)$  measures how far the cluster of $x$ reaches. We define  
\begin{equation*}
\Gn := \{X \in \X: \forall x \in X_{\La_n}: \rnx(x) \le |x| + 3 \log n\}
\end{equation*}
to be the set of all good configurations. 
\begin{lem} \label{legoodprop}
For all $X \in \Gn$ we have $m'(X) = m(X) + 1$, i.e. the proviso \eqref{altdefm} 
in the construction of $\Tn(X)$ never has to be used, for all $x \in X_{\Lan}$ 
we have $|\anx(x)| \le  |x| + 3 \log n + 2$ 
and in particular 
\begin{equation*} 
\forall x \in \La_{\sqrt n}: \tnx(x) = \de \ep \sqrt{\log n}. 
\end{equation*}
\end{lem}
This gives property (2). Our choice of $\ep$ implies that 
the corresponding continuum percolation is subcritical and thus large clusters 
have exponentially small probability, so it is reasonable to expect 
that good configurations in the above sense are not too rare. 
Indeed we can show property (4): 
\begin{lem} \label{legoodest}
For all $\bX \in \XX$ we have $\mu_n^z(\Gn^c|\bX) \le \frac 1 {n}$. 
\end{lem}
%

%========================================================================
%
%========================================================================

\newpage

\section{Proof of the lemmas from Subsection \ref{secproofthm}} 
\label{seclemmas}

\subsection{Basic properties: Lemmas \ref{lepropm} and \ref{leeasyprop}}

For the proof of Lemma \ref{lepropm} let $p \in \R^2$ and $t \ge 0$. 
From \eqref{deftn} it is easy to see that $\etn^0 = \tn(|.|)$ is Lipschitz-continuous 
with Lipschitz-constant $\frac{3 \de \ep}{\sqrt{\log n}} \frac 1 {n^{2/3}} \le \de$. 
By definition every function of the type $\mxt$ is continuous (wherever it is finite)
with a slope bounded by $\de$  (thanks to the proviso \ref{altdefm}).  
Since the minimum of two Lipschitz-continuous functions 
is again Lipschitz-continuous with the same Lipschitz-constant, 
it remains to be shown that 
$$
\forall x \in \R^2: |x-p|_2 = 1+\ep \Rightarrow \mxt(x) \ge \etn^0(x)
$$
in case of $\hxt \le \de \ep$. Indeed for $|x-p|_2 = 1+\ep$ by construction 
$$
\mxt(x) = t + \hxt = t + |\tn(|p| - 1 - \ep)-t| \ge \tn(|p| - 1 - \ep) \ge \tn(|x|) = \etn^0(x)
$$
since $|p|-1-\ep \le |x|$ and $\tn$ is decreasing. This finishes the proof 
of Lemma \ref{lepropm}. 

\bigskip

For the proof of Lemma \ref{leeasyprop} we use $\wedge$ as notation for the minimum. 
We consider the construction of $\Tn(X)$ for a fixed configuration $X \in \X$. 
For the first monotonicity property  \eqref{mono1} we note that for all $1 \le k \le m$ we have  
\[
\etaun^{k} \, = \, \etn^{k}(\epn^{k}) \, 
= \, \etn^{k-1}(\epn^{k}) \wedge m_{\epn^{k-1},\etaun^{k-1}}(\epn^{k}) \, 
\ge \, \etaun^{k-1},
\]
where the equalities follow from the recursive definition of $\etaun^k$ and $\etn^k$, 
and the last step from $\etn^{k-1}(\epn^k) \ge \etn^{k-1}(\epn^{k-1}) = \etaun^{k-1}$
by definition of $\epn^{k-1}$ and from $\mxt \ge t$. For the second monotonicity property  
\eqref{mono2} it suffices to note that $\etn^k$ is the minimum of $\etn^{k-1}$ 
and functions of the form $\mxt \ge t$, where $t \ge 0$. 

For \eqref{tnxalt} let $0 \le k \le l \le m$. The first equality is by definition, 
so it suffices to show that $\tn^k = \etn^l(\epn^k)$. We note that 
\begin{equation} \label{Klahc}
\alle x \in \R^2: \quad  
\etn^{l}(x) \, =\, \etn^{k}(x) \wedge \bigwedge_{k \le i <  l} \, 
 m_{\epn^i,\etaun^i}(x),
\end{equation}
so the above follows from $\etn^k(\epn^k) = \tn^k$, which is true by construction 
and  $m_{\epn^i,\etaun^i} \ge \etaun^i \ge \etaun^k$ for all $i \ge k$, 
which is true by \eqref{mono1}.

\eqref{tauab} is trivial in case 
of $x \notin \La_n$, since then all three terms equal $0$. In case of $x \in \Lan$ we have 
$x = \epn^k$ for some $k \ge 1$ and $\anx(\epn^k) = \epn^l$ for some $0 \le l \le k$. 
By definition the latter implies that $\epn^l \to \emptyset$, 
i.e. $\etn^0(\epn^l)= \etn^l(\epn^l) = \etaun^l$. Putting everything together an using 
\eqref{mono1} and \eqref{mono2} we see that $\etn(\epn^k) = \etaun^k$ satisfies 
$$
\tn(|\anx(\epn^k)|) = \etn^0(\epn^l) = \etaun^l \le \etaun^k
 = \etn^k(\epn^k) \le \etn^0(\epn^k), 
$$
which proves \eqref{tauab}. 
%
%For \eqref{proviso1} we note that $h_{\epn^{m'}, \etaun^{m'}} > \frac \ep 2$, which implies $m_{\epn^{m'}, \etaun^{m'}} = \etaun^{m'}$ is constant. Thus for 
%every $k > m'$ we have that 
%$$
%\etaun^{m'} \le  \etaun^k = \etn^k(\epn^k) \le m_{\epn^{m'}, \etaun^{m'}}(\epn^k) = \etaun^{m'},
%$$
%where in the first step we have used \eqref{mono1} and in the third step we have used 
%that $\etn^k \le m_{\epn^l,\etaun^l}$ for all $l < k$. This shows that 
%$\etaun^{k} = \etaun^{m'}$. On the other hand $\etn^{m'+1} \le \etaun^{m'}$ and 
%$m_{\epn^{l}, \etaun^{l}} \ge \etaun^l = \etaun^{m'}$ for every $l \ge m'$ imply that 
%$\etn^k = \etn^{m'+1}$ by definition of $\etn^k$. This finishes the proof of \eqref{proviso1}. 
\eqref{proviso2} is immediate from the construction: For $l < k$ such that 
$\epn^k \to \epn^l$ we have that $\etn^k(\epn^k) =  m_{\epn^l,\etaun^l}(\epn^k)$ 
is finite, which implies that $|\epn^k - \epn^l|_2 \le 1 + \ep$ as long as $l < m'$. 

The continuity property \eqref{Lcont} follows from Lemma~\ref{lepropm}, since
$\etn^k$ is the minimum of $\etn^0$ and functions of the form 
$m_{\epn^{l},\etaun^{l}}$, and since  
the minimum of Lipschitiz-continuous functions is again Lipschitz continuous. 

For the proof of \eqref{hc1} and \eqref{hc2} let $x_1,x_2 \in X$. 
Without loss of generality we may suppose that $x_1=\epn^l$ and $x_2=\epn^k$, 
where  $0 \le k < l$. 
If $|\epn^l - \epn^k|_2 \le 1$, then using \eqref{Klahc} for $x = \epn^l$ 
gives 
$\etaun^l = \etn^l(\epn^l) = \etaun^k$ 
since $\etn^k(\epn^l) \ge \etn^k(\epn^k) = \etaun^k$ and  
$m_{\epn^i,\etaun^i}(\epn^l) \ge \etaun^i \ge \etaun^k$ for all $k \le i < l$ 
(and equality in the case $i = k$). Thus we have shown \eqref{hc1}. 
Now we consider the case $|\epn^l - \epn^k|_2 > 1$. Let $x$ be the point of the line 
segment from $\epn^l$ to $\epn^k$ such that $|x - \epn^k|_2 = 1$. From \eqref{Klahc}
we get $\etn^l(x) \le \etaun^k$ since $m_{\epn^k,\etaun^k}(x) = \etaun^k$. 
Thus 
$$
0 \le \etaun^l - \etaun^k \le \etn^l(\epn^l) - \etn^l(x) \le 
\de |\epn^l - x|_2 = \de (|\epn^l - \epn^k|_2 - 1)
$$
by the Lipschitz-continuity of $\etn^l$ and choice of $x$. This implies 
\begin{align*}
&|\epn^l + \etaun^l \einh - \epn^k - \etaun^k \einh|_2
\ge |\epn^l- \epn^k|_2 -  |\etaun^l - \etaun^k|\\
&\ge |\epn^l- \epn^k|_2 - \de (|\epn^l - \epn^k|_2 - 1) 
= (1-\de) |\epn^l- \epn^k|_2 + \de > 1, 
\end{align*}
which proves \eqref{hc2} and thus finishes the proof of Lemma \ref{leeasyprop}.

%============================================================================

\subsection{Bijectivity of  the transformation: Lemma \ref{lebij}}
\label{secbijhc}

The proof of Lemma \ref{lebij} can be taken directly from \cite{R}. 
For sake of completeness we include a proof here. We have shortened and 
simplified some of the arguments.  
We aim to construct $\tTn$, the inverse transformation with respect to $\Tn$. 
In addition to the objects $\etn^k,\etaun^k$ and $\epn^k$ used in the construction 
of $\Tn$ we need the $k$-step transformation function 
$$
\eTn^k: \R^2 \to \R^2,\quad \eTn^k := id + \etn^k \cdot \einh. 
$$
By the Lipschitz-continuity of $\etn^k$, for every $c \in \R$ $\eTn^k(.,c)$ 
is continuous and strictly increasing, i.e. for every $0 \le k \le m$ we have 
\begin{equation}
\text{$\eTn^k$ is bijective and preserves the order of points 
on horizontal lines.} \label{propTn} 
\end{equation}
For reference we also note that for all $0 \le k \le m$ and $x \in \R^2$ we have 
\begin{equation} \label{inv}
(\eTn^k)^{-1}(x) + \etn^k((\eTn^k)^{-1}(x)) \einh = x.
\end{equation}
For defining an inverse transformation our main task is to reconstruct 
the enumeration of particles of a configuration when given only 
the transformed image of the configuration. 
The following lemma solves this reconstruction problem: 
\begin{lem} \label{lereconstruct} Let $X \in \X$, $\tX := \Tn(X)$ and 
$\etpn^k := \epn^k + \etaun^k \einh$ for $1 \le k \le m$. For every $k$ 
$\etpn^k$ is the point of $\tX_{\Lan} \weg \{\etpn^1, \ldots, \etpn^{k-1}\}$ 
at which the minimum of $\etn^k \circ (\eTn^k)^{-1}$ is attained. If there 
is more than one such point, then among those $\etpn^k$ is the smallest point 
with respect to lexicographic order.   
\end{lem} 
\Bew 
We first show that for all $1 \le k \le l$ 
\begin{equation} \label{inveincr}
\etn^{l} \circ (\eTn^{l})^{-1} \, \le \, \etn^k \circ (\eTn^k)^{-1}.
\end{equation}
For a proof let $x \in \R^2$, $x_k := (\eTn^k)^{-1}(x)$ and 
$x_{l} := (\eTn^{l})^{-1}(x)$. Both $x_k$ and $x_{l}$ are to the left of $x$. 
Since $\etn^{l} \le \etn^k$,  $\eTn^{l}(x_k)$ is left of $\eTn^k(x_k) = x$. 
By property \eqref{propTn} for $\eTn^{l}$ this implies that $x_{k}$ is 
left of $x_l$. Using \eqref{inv} this gives 
$\etn^{l}(x_{l}) = |x-x_{l}| \le |x-x_k| = \etn^k(x_k)$ and thus
proves \eqref{inveincr}. 
Now let $1 \le k \le l \le m$. By definition we have 
$\etn^{l}(\epn^l)  =  \etaun^l$, $\eTn^{l}(\epn^l) = \etpn^l$, 
$\etn^{k}(\epn^k)  =  \etaun^k$ and  $\eTn^{k}(\epn^k) = \etpn^k$. Using  
\eqref{mono2} and \eqref{inveincr} we deduce 
\[
 \etn^{k} (\eTn^{k})^{-1} (\etpn^k)\, 
= \, \etaun^k\, \le \, \etaun^l \, 
= \, \etn^{l} (\eTn^{l})^{-1} (\etpn^l) \,
\le \, \etn^{k} (\eTn^{k})^{-1} (\etpn^l).
\]
If $ \etn^{k} (\eTn^{k})^{-1} (\etpn^k) = \etn^{k} (\eTn^{k})^{-1} (\etpn^l)$, then 
all inequalities in the previous line have to be equalities, 
i.e. $\etaun^k = \etaun^l$ and $\etaun^l = \etn^{k} (\eTn^{k})^{-1} (\etpn^l)$. 
In light of \eqref{inv} this implies that $(\eTn^{k})^{-1} (\etpn^l) = \etpn^l - \etaun^l \einh  = \epn^l$, 
i.e. $\etn^k(\epn^l) = \etaun^l = \etaun^k$. 
Since lexicographic order is preserved by constant shifts, this 
concludes the proof of the lemma.  \qed 

\bigskip 

The above lemma motivates the following definition of $\tTn$. 
Let $\tX \in \X$ be an arbitrary configuration and $\tilde{m}(\tX)$ be the number of particles 
of $\tX_{\La_n}$. We recursively define an enumeration $\tpnx^k$ of the particles of $\tX_{\La_n}$, 
shift amounts $\ttaunx^k \in \R$ for the particles $\tpnx^k$, 
shift profiles $\ttnx^k: \R^2 \to [0,\infty)$ and corresponding transformations 
$\tTnx^k: \R^2 \to \R^2$. We then set 
\begin{align*}
&\tTn(X) \, := \, \{x - \ttnx(x) \einh : x \in X \}, \quad \text{ where }\\
&\ttnx(x) = 0 \text{ for } x \in \tX_{\La_n^c} \quad \text{ and } \quad 
\ttnx(\tpnx^k) = \ttaunx^k \text{ for } 1 \le k \le \tilde{m}(\tX). 
\end{align*}
In these notations we again omit the dependence on $\tX$ whenever it is 
clear which configuration $\tX$ is considered. For some fixed configuration $\tX \in \X$ 
we now describe the $k$-th step  $(1 \le k \le \tilde{m})$ of our recursive construction:  
\begin{itemize}
\item 
We set  $\ettn^k$ to be the minimum of $\ettn^{k-1}$ and the slow down 
$m_{\etpn^{k-1}- \ettaun^{k-1},\ettaun^{k-1}}$. Here $\ettn^0 = \tn(|.|)$ and 
$m_{\etpn^{0}-\ettaun^{0},\ettaun^{0}}$ is the minimum of all $m_{x,0}$ 
where $x \in \tX_{\La_n^c}$. 
\item 
Let $\etTn^k := id + \ettn^k \cdot \einh$. 
\item 
Let $\etpn^k$ be the point of $\tX_{\Lan} \weg \{\etpn^1, \ldots, \etpn^{k-1}\}$ 
at which the minimum of $\ettn^{k} \circ (\etTn^k)^{-1}$ is attained. 
If there is more than one such point then take the smallest point 
with respect to the lexicographic order. 
\item 
Let  $\ettaun^k := \ettn^{k}  \circ (\etTn^k)^{-1}(\etpn^k)$  be the corresponding minimal value. 
Here $\ettaun^0 := 0$. 
\end{itemize}

\bigskip 

We need to show that the above construction is well defined, i.e. 
that $\etTn^k$ is invertible in every step, and has suitable 
monotonicity properties. 
\begin{lem} \label{leeiginv}
Let $\tX \in \X$ and $k \ge 0$. Then
\begin{align}
& \etTn^k \text{ is bijective and preserves the order of points on horizontal lines}, \label{lipinv}\\
%& (\etTn^k)^{-1} + \ettn^k \circ (\etTn^k)^{-1} \einh  \, = \,id, 
% \label{TSinv}\\
%& \alle c \in \R ,x \in \R^2: \; \ettn^k \circ (\etTn^k)^{-1}(x) 
% \,\ge\,c\, 
%  \Leftrightarrow \, \ettn^k(x -  c\einh) \, \ge \, c, \label{tsinv}\\ 
&0 = \ettaun^0 \le \ettaun^1 \le \ldots \le \ettaun^{m}, \label{monoinv1}\\
&\ettn^0 \ge \ettn^1 \ge ... \ge \ettn^m \ge 0. \label{monoinv2}
\end{align} 
\end{lem}
\Bew 
Whenever $\ettn^k$ is well defined, it satisfies the Lipschitiz property 
analogous to \eqref{Lcont}, which gives \eqref{lipinv}. Inductively one
can thus show that the above construction is well defined. 
\eqref{monoinv2} can be show similarly to the corresponding monotonicity 
property \eqref{mono2}. 
For \eqref{monoinv1} we will show
$\ettaun^{l} \ge \ettaun^{k}$ for every $0 \le k < l$. 
Let $x := \etpn^{l}$, $x_{k} = (\etTn^{k})^{-1}(x)$, 
$x_{l} = (\etTn^{l})^{-1}(x)$ and $x_{k}' := x - \ettaun^{k} \einh$. 
All $x_{l},x_{k}$ and $x_{k}'$ are to the left of $x$. 
Since $\ettn^{k}(\etTn^{k})^{-1}(x) \ge \ettaun^{k}$ by definition of $\etpn^k$, 
and since we have \eqref{inv} for $\ettn^{k}$, $x_{k}$ is to the left of $x_{k}'$. 
By property \eqref{lipinv} for 
$\etTn^{k}$ this implies that $\etTn^{k}(x_{k}')$ is to the right 
of $\etTn^{k}(x_{k}) = x$. Since for $\etTn^{l}$ the shift 
is slowed down at most to the value $\ettaun^{k}$ as compared to 
$\etTn^{k}$ and since $|x-x_{k}'| = \ettaun^{k}$, 
we still have that $\etTn^{l}(x_{k}')$ is to the right 
of $x$. By \eqref{lipinv} for $\etTn^{l}$ thus $x_{l}$ is to the left of $x_{k}'$, 
and thus $\ettaun^{l} = |x-x_{l}| \ge |x-x_{k}'| = \ettaun^{k}$. 
\qed

\bigskip

In order to show that  $\tTn$ really is the inverse of $\Tn$ we need an 
analogue of the reconstruction result from Lemma \ref{lereconstruct}.
\begin{lem} \label{lereconstructinv}
Let  $\tX \in \X$, $\ettn^k$, $\etTn^k$, $\etpn^k$ and  $\ettaun^k$ 
as above, $X := \tTn(\tX)$ and 
$\epn^k := \etpn^k - \ettaun^k \einh$. 
For every $1 \le k \le \tilde{m}$ $\epn^k$ is the point of 
$X_{\Lan} \weg \{\epn^1, \ldots, \epn^{k-1}\}$ at which the minimum of 
$\ettn^k$ is attained. If there is more than one such point, then among those
$\epn^k$ is the smallest point with respect to lexicographic order. 
\end{lem} 
\Bew 
Let $1 \le k \le l \le \tilde{m}$. By \eqref{inv} for $\etTn^{k}$ we have  
$(\etTn^{k})^{-1}(\etpn^k) = \etpn^k - \ettaun^k \einh = \epn^k$, which implies 
$\ettn^{k} (\epn^k) =  \ettaun^k$.
Similarly $\ettn^{l} (\epn^l) =  \ettaun^l$. Using \eqref{monoinv1} and 
\eqref{monoinv2} we obtain  
\[
\ettn^{k} (\epn^k) =   \ettaun^k \le  \ettaun^l  
=  \ettn^{l} (\epn^l)  \le  \ettn^{k} (\epn^l).
\]
If $\ettn^{k} (\epn^k) = \ettn^{k} (\epn^l)$, then all inequalities in the previous line
have to be equalities, so $ \ettaun^k =  \ettaun^l$
and $\ettn^{l} (\epn^l)  = \ettn^{k} (\epn^l)$. This implies 
$\etpn^l = \etTn^{l} (\epn^l)  = \etTn^{k} (\epn^l)$, i.e. $(\etTn^k)^{-1}(\etpn^l) = \epn^l$, 
which gives $\ettn^k((\etTn^k)^{-1}(\etpn^l)) = \ettaun^l =  \ettaun^k$ 
in light of \eqref{inv}. 
Since lexicographic order is preserved by constant shifts, this concludes the proof of 
the lemma. 
\qed 
\begin{lem}\label{leinvhc}
On $\X$ we have $ \quad  \tTn \circ \Tn \, = \, id \quad $ 
and  $ \quad  \Tn \circ \tTn \, = \, id$.
\end{lem}
\Bew
For the first part let  $X \in \X$ and $\tX := \Tn(X)$. We have $X_{\Lan^c} = \tX_{\Lan^c}$ by construction
and $\tilde{m}(\tX) = m(X)$ by \eqref{propTn}. Now it suffices to prove 
\begin{equation}  \label{tildegleichhc}
\ttnx^k \, = \, \tnx^k, \; \tTnxn^k \, = \, \Tnxn^k, \; 
\ttaunx^k \, = \, \taunx^k \, \text{ and } \, 
\tpnx^k \, = \, \pnx^k +  \taunx^k  
\end{equation} 
for every $k \ge 0$ by induction on  $k$. Here $\tpnx^0 = \pnx^0 +  \taunx^0$
is interpreted as $X_{\Lan^c} = \tX_{\Lan^c}$. The case $k=0$ is trivial. 
For the inductive step  $k-1 \to k$ we observe that  
$\ettn^k = \etn^k$ by induction hypothesis, and $\etTn^k  =  \eTn^k$ 
is an immediate consequence. Combining this with Lemma \ref{lereconstruct}
and the definition of $\etpn^k$ we get 
$\etpn^k = \epn^k  +  \etaun^k$ and   $\ettaun^k = \etaun^k$.\\
For the second part let  $\tX \in \X$ and 
$X := \tTn(\tX)$. As above it suffices to show \eqref{tildegleichhc}
by induction on $k$. Here the inductive step follows from Lemma  
\ref{lereconstructinv}.
\qed
  
%===========================================================================

\newpage 

\subsection{Density of the transformed process: Lemma \ref{ledensphc}}
\label{secledensphc}

Again, the proof of Lemma \ref{ledensphc} can be taken directly from \cite{R}. 
For sake of completeness we include a proof here. We have shortened and 
simplified some of the arguments.  
Let $\bX \in \XX$ and $f \ge 0$ be measurable. By definition of $\mu_n^z(.|\bX)$ 
we have  
\begin{align*}
&\int d\mu_n^z(dX|\bX) \, f(\Tn(X))\ph(X)\\
&= \frac 1 {Z^z_n(\bX)} e^{-(2n)^2} \sum_{m \ge 0} z^m 
\int_{{\Lan}^{m}} dx f(\Tn(\bX_x)) \ph(\bX_x) 1_{\XX}(\bX_x)
\end{align*}
using the shorthand notation $\bX_x := \{ x_i: i \in J\} \cup \bX_{\Lan^c}$
for  $x  \in \Lan^{\;J}$.
By \eqref{hc1} and \eqref{hc2} we have that $\bX_x \in \XX$ iff 
$\Tn(\bX_x) \in \XX$ and thus $f(\Tn(\bX_x))1_{\XX}(\bX_x) = 
(f 1_{\XX}) (\Tn(\bX_x))$. Incorporating $1_{\XX}$ into $f$ 
it thus suffices to show that 
$$
\int_{{\Lan}^{m}} dx f(\Tn(\bX_x)) \ph(\bX_x) = 
 \int_{{\Lan}^m} dx' f(\bX_{x'}) \quad \text{ for all } m \ge 0. 
$$
 Since we aim at using the Lebesgue transformation 
theorem, we would like to enumerate the particles of $\bX_x$, preferably 
in the same order as they occur in the construction of $\Tn(\bX_x)$. 
So let $\Pi$ be the set of all permutations 
$\per:\{1,\ldots ,m\}\to \{1,\ldots,m\}$. For $\per \in \Pi$ 
let 
\[ \begin{split}
\ak \, &:= \, \big\{ x \in\Lan^{\;m}: \alle 1 \le k \le m: 
 x_{\per(k)} = P^k_{n,\bX_x}\big\}\quad \text{ and } \\
\tak \, &:= \, \big\{ x \in\Lan^{\;m}: \alle 1 \le k \le m:
 x_{\per(k)} = \tilde{P}^k_{n,\bX_x}\big\},
\end{split}\]
where  $\tilde{P}^k_{n,\bX_x}$ are the points from the construction 
of the inverse transformation in Subsection \ref{secbijhc}.  
Both $\ak$ and $\tak$ form a disjoint decomposition of $\Lan^m$, so it suffices 
to show that for all $\eta \in \Pi$ we have. 
$$
\int dx 1_{\ak}(x)  f(\Tn(\bX_x)) \ph(\bX_x) = 
 \int dx' 1_{\tak}(x') f(\bX_{x'}).   
$$
By reordering the components of $x$ and $x'$ according to $\eta$ 
(and using that a product measure is invariant under permutation of components) 
it suffices to show the above equality for $\eta = id$. We simplify our notation 
by setting $A := A_{id}$ and $\tilde{A} := \tilde{A}_{id}$. 
We now try to express $\Tn(\bX_x)$ as a corresponding transformation $T(x)$. 
For $x \in \Lan^{\;k}$ we define a formal transformation 
$T(x)  :=  (T^k_{x} (x_k))_{1 \le k \le m}$ recursively by 
$$
t^k_{x} := t^{k-1}_{x} \wedge m_{x_{k-1},\tau^{k-1}_{x}},\quad 
\tau^{k}_{x} := t^{k}_{x}(x_k), \quad  
T^{k}_{x} = id + t^{k}_{x} \cdot \einh, 
$$
where $t^{0}_{x} = \etn^0$ and 
$m_{x_{0},\tau^{0}_{x}}$ is the minimum of all functions
$m_{x,0}$ where $x \in \bX_{\Lan^c}$. 
By definition for $x \in A$ we have $x_k = P^k_{n,\bX_x}$ for all $1 \le k \le m$, which 
inductively implies that
\begin{equation} \label{at1}
t^k_{n,\bX_x} \, = \, t^k_{x}, \quad   
T^k_{n,\bX_x} \, = \, T^k_{x} \quad \text{ and } \quad  
\tau^k_{n,\bX_x} \, = \, \tau^k_{x}
\end{equation}
for all $1 \le k \le m$ and thus 
\begin{equation} \label{at2}
 \T_n(\bX_x) \, = \, \bX_{T(x)}.
\end{equation}
We also note that by construction for every $1 \le k \le m$ 
\begin{equation} \label{dep}
\text{ both  $t^k_{x}$ and $T^k_{x}$ depend on $x$ 
only through $x_1,...,x_{k-1}$. }
\end{equation}
Furthermore we observe that for all  $x \in (\R^2)^{k}$  we have 
\begin{equation} \label{trapo}  
x \in A \quad \Leftrightarrow \quad T(x) \in \tilde{A}.
\end{equation}
Here ``$\Rightarrow $'' holds by \eqref{at2} and \eqref{tildegleichhc} 
from the proof of Lemma  \ref{leinvhc}. For  ``$\Leftarrow $''
let  $x \in  (\R^2)^{k}$ such that $T(x) \in \tilde{A}$ and
let $X':= \tTn(\bX_{T(x)})$, 
where $\tTn$  is the inverse of  $\Tn$ as defined in the last subsection. 
By induction  
\[
\alle 1 \le k \le m:  \quad T_{n,X'}^k =  T^k_{x} \quad \text{ and } \quad  
x_{k} = P_{n,X'}^k.
\]
In the inductive step $k-1 \to k$ the first assertion follows from the
induction hypothesis and the second 
follows from the bijectivity of $T_{n,X'}^k$ and  
\[
T_{n,X'}^k(x_k) \, = \,  T^k_{x}(x_k) \, 
= \, \tilde{P}^k_{n,\bX_{T(x)}} \, 
= \, P^k_{n,X'} + \tau^k_{n,X'} \, = \, T_{n,X'}^k(P_{n,X'}^k), 
\]
which follows from $T_{n,X'}^k =  T^{k}_{x}$, the definition of $\tilde{A}$ 
and \eqref{tildegleichhc} from the proof of Lemma \ref{leinvhc}. 
This completes the proof of the above assertion 
and we conclude $\bX_x= X'$, which implies 
$x_k  = P_{n,X'}^k = P^k_{n,\bX_x}$. Thus \eqref{trapo} holds. 
We now get 
\begin{align*}
\int dx &1_A(x)  f(\Tn(\bX_x)) \ph(\bX_x) =
\int dx 1_A(x) f(\bX_{T(x)}) \prod_{k=1}^m  
\big|1 + \partial_{\einh} t^{k}_{x}(x_k)\big| \\
&= \Big[ \prod_{k=1}^{m}  \int dx_k \,
 \big|1 + \partial_{\einh} t^{k}_{x}(x_k)\big| \Big] \, 
 g(T(x)),  
\end{align*}
where we have used the definition of $\ph$, \eqref{at1} and \eqref{at2} and 
finally \eqref{trapo} using the shorthand notation 
$g: (\R^2)^{k} \to \R$, $g(x) :=  1_{\tilde{A}}(x) f(\bX_x)$.
Now we transform the integrals. For $k=m$ to $1$ we substitute 
$x_k' := T^k_{x}(x_{k})$, making use of \eqref{dep}. 
As before it can be seen that $t^{k}_{x}$ is Lipschitz-continuous 
with Lipschitz-constant $ \le \de$, so  
$T^{k}_{x}$ is bijective and Lipschitz-continuous and by Rademacher's theorem 
thus differentiable almost everywhere. Indeed we have 
$$
\nabla T^{k}_{x} = \begin{pmatrix}
 1 + \partial_{\einh} t^{k}_{x} \quad &   ... \\
 0 & 1
\end{pmatrix} \quad \text{ and thus } \quad 
dx_k' \, = \,  dx_k \big| 1 + \partial_{\einh} t^{k}_{x}(x_k)\big|
$$
for all $k$ by the Lebesgue transformation theorem. 
So the above integral reduces to 
$$
\Big[\prod_{k=1}^{m}  \int dx'_k \Big] \, g(x') \, 
=  \, \int dx' \, 1_{\tilde{A}}(x') f(\bX_{x'}),
$$
which finishes the proof of Lemma \ref{ledensphc}.

%=======================================================================

\subsection{Estimate of the shift amount: Lemma~\ref{legoodprop}}

Let $X \in \Gn$. We will first show that $h_{\epn^k,\etaun^k} \le \de \ep$ 
for all $0 \le k \le m$ 
by induction on $k$. For the case $k = 0$ we observe that for every $x \in X_{\Lan^c}$ 
we have 
$$
h_{x,0} = \tn(|x| -1-\ep) \le \tn(n-2) \le \frac {3 \de \ep}{\sqrt{\log n}}
\log\frac{n}{n-2} \le \de \ep.  
$$
In the inductive step we may assume that 
$h_{\epn^i,\etaun^i} \le \de \ep$ for all $0 \le i < k$, which implies that 
$m' \ge k$, and using \eqref{proviso2} we obtain $|\anx(\epn^k)| \le \rnx(\epn^k)+1 + \ep
\le |\epn^k| + 3 \log n + 1 + \ep$, where we have also used $X \in \Gn$.  
We thus get 
\begin{align*}
&h_{\epn^k,\etaun^k} = \tn(|\epn^k|-1-\ep) - \etaun^k \le 
\tn(|\epn^k|-1 - \ep) - \tn(|\anx(\epn^k)|) \\
&\le \frac{3 \de \ep}{\sqrt{\log n} n^{2/3}} (|\anx(\epn^k)| - |\epn^k| + 1 + \ep)
\le \frac{3 \de \ep}{\sqrt{\log n} n^{2/3}} (3 \log n +2.5) \le \de \ep. 
\end{align*}
Here we have used the monotonicity of $\tn$, the estimate \eqref{tauab} for $\etaun^k$, 
the upper bound $\frac{3 \de \ep}{\sqrt{\log n} n^{2/3}}$ on the derivative of $\tn$ and $n \ge 200$.  
This finishes the induction step. 
In the inductive step we have also shown that 
$|\anx(x)| \le |x| + 3 \log n + 2$. 
For $x \in X_{\La_{\sqrt n}}$ this is bounded by $n^{2/3}$, so 
\eqref{tauab} implies that $\etn(x) = \de \ep \sqrt{\log n}$. 
This finishes the proof of Lemma \ref{legoodprop}.

%==========================================================================

\subsection{Strategies for estimating probabilities}

In the following sections we need to estimate expectations of sums such as 
$\sum_{x \in X} f(x)$ with respect to $\mu_n^z(.|\bX)$. We present two different 
strategies for such estimates. The first one relies on the hard core of particles, 
the second one  on properties of the underlying Poisson point process.
\begin{lem} \label{leprobabhc}
Let $X \in \XX$ and $f,g \ge 0$ be measurable functions  on $\R^2$ such that 
$f(y) \ge g(x)$ for all $y \in B_x := \{x' \in \R^2: |x'-x|_2 < 1/2\}$. 
Then 
\begin{equation} \label{probabhc}
\sum_{x \in X} g(x) \le \frac 4 \pi \int dx f(x). 
\end{equation}
\end{lem}

\Bew We have $g(x) \le \frac 4 \pi \int_{B_x} dy f(y)$, so 
$$
\sum_{x \in X} g(x) \le \sum_{x \in X} \frac 4 \pi \int_{B_x} dy f(y) \le \frac 4 \pi \int dy f(y). 
$$
In the last step we have used that the disks $B_x, x \in X$ are disjoint because of $X \in \XX$. 
\qed 

\bigskip 

\begin{lem} \label{leprobab}
let $\bX \in \XX$, $z > 0$ and $\La \in \B^2$ bounded. Let $g\ge 0$ be measurable on $\La \times \X$. 
We have 
\begin{equation} \label{probab}
\int \mu^z_\La(dX|\bX) \sum_{x \in X_{\La}}  g(x,X)
\le \, z \int_{\La} dx \int \mu^z_\La(dX|\bX)  g(x,X \cup\{x\}).
\end{equation}
\end{lem}

\Bew \eqref{probab} relies on a corresponding property of the Poisson point process:  
For  $\bX, z, \La$ as above and $f \ge 0$ measurable on $\La \times \X$ we have 
\begin{equation} \label{probabP}
\int \nu_{\La}(dX|\bX) \sum_{x \in X_{\La}} f(x,X)  
=  \int \nu_{\La}(dX|\bX) \int_{\La}   dx f(x,X \cup\{x\}).
\end{equation}
To show this, we note that by definition of $\nu_{\La}(.|\bX)$ the left hand side equals 
$$
e^{-\la^2(\La)} \sum_{k \ge 0} \frac 1 {k!} \int_{\La} dx_1 ... \int_{\La} dx_k
\sum_{1 \le l \le k} f(x_l,\{x_1,...,x_k\} \cup Y_{\La^c}).
$$
Since the product measure is invariant under permutations, the above equals 
$$
e^{-\la^2(\La)} \sum_{k \ge 1} \frac k {k!}   \int_{\La} dx_1 ... \int_{\La} dx_k
f(x_k,\{x_1,...,x_k\} \cup Y_{\La^c}).
$$
Substituting $k' := k-1$, $x := x_k$ and $X = \{x_1,...,x_{k-1}\}\cup Y_{\La^c}$
the definition of the Poisson point process implies that the above expression equals the right hand side of 
\eqref{probabP}. This finishes the poof of \eqref{probabP}. Applying \eqref{probabP} to the function 
$f(x,X) := g(x,X)  1_{\XX}(X)  z^{\# X_\La}$ and dividing by $Z_\La^z(\bX)$ we obtain 
\begin{equation*}
\int \mu^z_\La(dX|\bX)  \sum_{x \in X_{\La}} \! g(x,X)= \, \int \mu^z_\La(dX|\bX) \int_{\La} \!  dx
g(x,X \cup\{x\})  z 1_{\{...\}},
\end{equation*}
where the indicator enforces that $x$ keeps hard core distance from the particles of $X$. 
Estimating this indicator by $1$ finishes the proof of \eqref{probab}. \qed 

\bigskip 

Multiple sums can be treated by applying the above estimates iteratively. 
Using $\Sigma^{\neq}$ as a shorthand notation for a multiple sum such that
the summation indices are assumed to be pairwise distinct we thus get the following: 

\begin{lem} \label{leprobabmult}
let $\bX \in \XX$, $z > 0$ and $\La \in \B^2$ bounded. Let $f \ge 0$ be measurable on $\La^{m+1}$, 
where $m \ge 1$. 
We have 
\begin{equation}\begin{split} \label{probabmult}
&\int \mu^z_\La(dX|\bX)   \!\! \sideset{}{^{\neq}} \sum_{x_1,\ldots,x_{m} \in X_{\La}, x_0 \in X} \!\!\!
f(x_0,\ldots,x_{m})\\ 
&\le \, z^m \!\! \int \mu^z_\La(dX|\bX)  \sum_{x_0 \in X} \int_{\La} \!\!  dx_1 ...  \int_\La \!\! dx_{m}  
f(x_0,\ldots,x_{m}).
\end{split}\end{equation}
\end{lem}
\Bew Applying \eqref{probab} to $g(x_m,X) := \sideset{}{^{\neq}} \sum f(x_0,x_1,\ldots,x_m)$, 
where the sum is over all $x_1,\ldots,x_{m-1} \in (X \setminus \{x_m\})_{\La}$ and 
all $x_0 \in X \setminus \{x_m\}$, we obtain 
\begin{align*}
\int \mu^z_\La(dX|\bX)  &\sideset{}{^{\neq}} \sum_{x_1,\ldots,x_m \in X_{\La},x_0 \in X} \!\!
f(x_0,\ldots,x_m)\\ 
&\le \, z \!\! \int_{\La} \!\!  dx_m \int \mu^z_\La(dX|\bX) \!\!
\sideset{}{^{\neq}} \sum_{x_1,\ldots,x_{m-1} \in X_{\La},x_0 \in X} f(x_0,\ldots,x_m).
\end{align*}
We note that on the right hand side we have replaced the sum over 
$x_0 \in (X \cup \{x_m\}) \setminus \{x_m\}$ by a sum over $x_0 \in X$, which doesn't affect the value of 
the right hand side (and similarly for the sum over $x_1,..,x_{m-1}$). 
Proceeding inductively we obtain \eqref{probabmult}.  \qed

\bigskip

%===============================================================================================

\subsection{Estimate of the cluster size: Lemma \ref{legoodest}}

Let $X \in \XX$. If $X$ is bad, there is a $x \in X_{\Lan}$ such that $\rnx(x) > |x| + 3 \log n$, 
i.e. there are distinct $x_0, x_1,...,x_N \in X_{\La_n}$
such that $x_0 \sim x_1 \sim ... \sim x_N$ and $N \ge \frac{3}{ 1+\ep} \log n$. Fixing $N := \lceil \frac{3}{ 1+\ep} \log n \rceil$ and introducing the notation 
$$
A_\ep(x) := \{y \in \R^2: 1 \le |y-x|_2 \le 1+\ep\}
$$ 
for an annulus centred at  $x$, the above implies that $x_{i+1} \in A_\ep(x_i)$ for all $i$, and so  
\begin{align*}
&\mu_n^z(\Gn^c|\bX) \le \int \mu_n^z(dX|\bX) \sideset{}{^{\neq}} \sum_{x_0,\ldots ,x_N \in X_{\Lan}}   
 1_{\{\forall i: x_{i+1} \in A_\ep(x_i) \}} \\
&\le z^N \!\! \int \mu_n^z(dX|\bX) \sum_{x_0 \in X_{\La_n}} \int_{\La_n} \!\!  dx_1 ...  \int_{\La_n} \!\! dx_N 
 1_{\{\forall i: x_{i+1} \in A_\ep(x_i) \}} \\
&= z^{N} \int \mu_n^z(dX|\bX) \sum_{x_0 \in X_{\La_n}}  \int_{A_\ep(x_0)} \!\!  dx_1 ...  
\int_{A_\ep(x_{N-1})} \!\! dx_N .
\end{align*}
Here we have used \eqref{probabmult}. Estimating the integrals using 
$$\la^2(A_\ep(x))  = \pi(2 \ep + \ep^2)  \le \frac {9\pi} 4 \ep \le 8 \ep,
$$ 
we obtain 
\begin{align*}
&\mu_n^z(\Gn^c|\bX) \le 
\int \mu_n^z(dX|\bX) \sum_{x_0 \in X}  1_{\La_n}(x_0) (8 \ep z)^N \\
&\le (8 \ep z)^N  \frac 4 \pi \int dx_0 1_{\La_{n+0.5}}(x_0)  
\le \frac 1 {6^N} \frac 4 \pi (2n+1)^2 \le \frac 4 {\pi} \frac {(2n+1)^2} {6 n^3} \le \frac 1 n.  
\end{align*}
Here we have used \eqref{probabhc}, the definition of $\ep$ and  the 
estimate $6^N \ge 6 n^3$.

%==========================================================================

\subsection{Estimation of the densities: Lemma \ref{ledensest}}

Let $X \in \XX$. We first note that 
$$
\ph(X)\bph(X) =  \prod_{k = 1}^{m(X)} \big|1 +  \partial_{\einh} \tnx^{k} (\pnx^k)\big|
\cdot \big|1 -  \partial_{\einh} \tnx^{k} (\pnx^k)\big|.
$$
By the Lipschitz-continuity from \eqref{Lcont} we have 
$|\partial_{\einh} \tnx^k(\pnx^k)| \le 1/2$. 
Using $|\log(1-a)| \le  \frac 4 3 a$ for $0 \le a \le 1/4$ we obtain
$$
|\log ( \ph(X) \bph(X))| \;
= \, \big| \sum_{1 \le k \le m} 
 \log\big(1 - (\partial_{\einh} \etn^k(\epn^k))^2\big)\big| 
 \le \, \sum_{1 \le k \le m} \frac 4 3  (\partial_{\einh} \etn^k(\epn^k))^2.
$$
By construction of $\etn^k$ its derivative either equals the derivative of $\etn^0$ 
or of one of the functions $m_{\epn^{l},\etaun^{l}}$ such that $0 \le l< k$, 
whenever it exists. Furthermore in case of $k > m'$ its derivative equals $0$, since 
in this case the proviso \eqref{altdefm} implies that $\etn^k(x) \le \etn^k(\epn^k) = \etaun^{m'}$ for all $x$. 
Thus the above can be estimated by the sum of the two following terms: 
\begin{align*} 
&\Sigma_1^n(X) := \sum_{1 \le k \le m} \!\!  \frac 4 3 (\partial_{\einh} \etn^0(\epn^k))^2, \quad 
\Sigma_2^n(X) := \sum_{1 \le k \le m'} \sum_{0 \le l < k}  \!\! \frac 4 3 (\partial_{\einh} m_{\epn^{l},\etaun^{l}}(\epn^k))^2 
\end{align*} 
For estimating $\Sigma_1^n$ we use \eqref{probabhc} to obtain 
\begin{align*}
\Sigma_1^n(X) &\le \frac 4 3  \sum_{x \in X_{\La_n}} (\partial_{\einh} \etn^0(x))^2
\le \frac{12 \de^2 \ep^2 }{\log n } \sum_{x \in X} \frac{1_{[n^{2/3}, n]}(|x|)}{|x|^2}\\ 
&\le \frac{48 \de^2 \ep^2 }{\pi \log n } \int  \!\! dx \frac {1_{[n^{2/3}-0.5, n+0.5]}(|x|)} {|x-0.5|^2}
\le \frac{3 \de^2 }{\pi \log n } 3 \log n \le 3 \de^2.
\end{align*}
Here we have used $\ep \le \frac 1 4$ and substituted $s := |x|$ for calculating 
the integral:  
\begin{align*}
&\int \! dx \frac {1_{[n^{2/3}-0.5, n+0.5]}(|x|)} {(|x|-0.5)^2}
\le \int_{n^{2/3} - 0.5}^{n+0.5} ds \frac {8s}{(s- 0.5)^2}    
\le \int_{n^{2/3} - 1}^{n} dt \frac {8t+4}{t^2}\\
&\le 8 \log \frac{n}{n^{2/3} -1} + \frac 4 {n^{2/3} - 1}
= \frac 8 3 \log n + 8 \log  \frac{n^{2/3}}{n^{2/3} -1} + \frac 4 {n^{2/3} - 1} 
\le 3 \log n
\end{align*}
using $n \ge 200$ in the last step. For $\Sigma_2^n$ we estimate $|\partial_{\einh} m_{\epn^l,\etaun^l}(\epn^k)|$ by 
\begin{align*}
& \frac 1 \ep \big(\tn(|\epn^{l}|-1-\ep) - \etaun^{l}\big) \le \, \frac 1 \ep \big(\tn(|\epn^{l}|-1-\ep) - \tn(|\anx(\epn^{l})|)\big) \\
&
\le \frac{3 \de \ep  c(|\epn^l|)}{\ep \sqrt{\log n}}(|\anx(\epn^{l})| - |\epn^{l}|+1+\ep)), 
\end{align*}
using \eqref{tauab} in the first step and estimating the slope of  $\tn$ 
in the second step  setting 
$$
c(s) := \frac{1}{\max\{s-1-\ep, n^{2/3} \}}. 
$$
We set $k = k_0$, $l = k_1$ and 
note that we have  $m' \ge k_0 > k_1 >  ... > k_N \ge 0$ (with $N \ge 1$) such that 
for $x_i := \epn^{k_i}$ we have $x_0 \to x_1 \to .. \to x_N = \anx (x_1) \to \emptyset$. 
In particular $x_0,...,x_N$ are distinct, $x_0,...,x_{N-1} \in \Lan$ and 
by \eqref{proviso2} we have $|x_i - x_{i+1}|_2 \le 1 + \ep$ for all $i$. 
This gives $|x_N| - |x_1| \le |x_N-x_1|_2 \le (1+\ep)(N-1)$. 
Treating the cases $N = 1$ and $N \ge 2$ separately, the above implies that $\Sigma_2^n(X)$ 
can be estimated by the sum of 
\begin{align*}
\Sigma_{2,1}^n(X) &:=  \sideset{}{^{\neq}} \sum_{x_0 \in X_{\Lan}, x_1 \in X} 
\frac{12 \de^2(1+\ep)^2}{\log n} c(|x_1|)^2  1_{A_\ep(x_1)}(x_0) \quad \text{ and } \\
\Sigma_{2,2}^n(X) &:= \sum_{N \ge 2} \quad \sideset{}{^{\neq}} \sum_{x_0,...,x_{N-1} \in X_{\Lan}} 
\frac{12 \de^2(1+\ep)^2N^2}{\log n} c(|x_1|)^2  1_{\{\forall i: x_i \in A_\ep(x_{i-1})\}}.
\end{align*}
Using \eqref{probabmult} we get 
\begin{align*}
\mu_n^z(\Sigma_{2,1}^n|\bX) \le 
&\frac{12(1+\ep)^2\de^2z}{\log n} \int \mu_n^z(dX|\bX)  \int_{\Lan} dx_0 
\sum_{x_1 \in X_{\La_{n+1+\ep}}}
\!\!\! c(|x_1|)^2  1_{A_\ep(x_1)}(x_0).
\end{align*}
Now we first estimate $\int_{\La_n} dx_0  1_{A_\ep(x_1)}(x_0) \le  \la^2(A_\ep(x_1)) \le 8 \ep$, 
and then use \eqref{probabhc} to estimate the sum over $x_1$. For this we note that  
\begin{align*}
& \int_{\La_{n+1.5+\ep}} \!\!\!\!\!\!\! dx_1 c(|x_1| - 0.5)^2
\le \int_0^{n+1.5+\ep} \!\!\! ds \frac{8s}{\max\{(s-1.5-\ep)^2,n^{4/3} \}}\\
&=  \int_0^{n^{2/3}+1.5+\ep} \!\!\! ds \frac{8s}{n^{4/3}} + \int_{n^{2/3}+1.5+\ep}^{n+1.5+\ep} \!\!\!  ds \frac{8s}{(s-1.5-\ep)^2}\\
&\le \frac 4 {n^{4/3}} (n^{2/3}+2)^2 + \int_{n^{2/3}}^{n} dt \frac{8(t+2)}{t^2}
\le 4 (1 + \frac{2}{n^{2/3}})^2 +  \int_{n^{2/3}}^{n} dt \frac{8}{t} + \int_{n^{2/3}}^{\infty} dt \frac{16}{t^2}\\
&\le 5 + \frac 8 3  \log n \le 4 \log n. 
\end{align*}
Here we have set $s := |x_1|$ and $t := s-1.5-\ep$ and used that $n \ge 200$. 
Thus we finally obtain 
$$
\mu_n^z(\Sigma_{2,1}^n|\bX) \le 
\frac{12(1+\ep)^2\de^2z}{\log n} \cdot 8 \ep \cdot \frac 4 {\pi} 4 \log n
\le \frac{192(8 \ep z)(1+\ep)^2\de^2 }{\pi} \le 16 \de^2.
$$
The expectation $\mu_n^z(\Sigma_{2,2}^n|\bX)$ can be estimated similarly. We note that
\begin{align*}
&\int \mu_n^z(dX|\bX) \sideset{}{^{\neq}} \sum_{x_0,...,x_{N-1} \in X_{\Lan}} 
 c(|x_1|)^2  1_{\{\forall i: x_i \in A_\ep(x_{i-1})\}}\\
&\le  \int \mu_n^z(dX|\bX)  z^{N-1} \sum_{x_1 \in X_{\Lan}} c(|x_1|)^2 ...  \int_{A_\ep(x_{N-2})} \!\!\! dx_{N-1}\int_{A_\ep(x_1)} \!\!\! dx_0    \\
&\le  \int \mu_n^z(dX|\bX) \sum_{x_1 \in X_{\Lan}} c(|x_1|)^2 (8 \ep z)^{N-1}
\le  \frac 4 {\pi} \int_{\La_{n+0.5}} dx_1 c(|x_1|-0.5)^2 (\frac 1 6)^{N-1}\\
&\le \frac {16 \log n} {\pi} (\frac 1 6)^{N-1},
\end{align*}
first using \eqref{probabmult} and estimating the arising integrals, then estimating the sum over $x_1$
using  \eqref{probabhc} and estimating the arising integral as above. 
Thus 
$$
\mu_n^z(\Sigma_{2,2}^n|\bX) \le 
 \frac{192 \de^2(1+\ep)^2}{\pi} \sum_{N \ge 2} \frac{N^2}{6^{N-1}} 
 \le 100  \de^2
$$
Putting everything together we get 
$$
\mu_n^z( |\log ( \ph \bph)| | \bX) = (2+ 16 + 100) \de^2 \le 120 \de^2.
$$

\end{sloppypar}

% %===========================================================================

\renewcommand{\thesection}{}

\setlength{\parindent}{0cm}

\end{document}